\documentclass[twocolumn]{aastex631}

\usepackage{enumitem}
\usepackage{booktabs}
\usepackage{graphicx}
\usepackage{amsmath}
\usepackage[nolist]{acronym}
\usepackage[nameinlink,capitalise,noabbrev]{cleveref}
\crefname{deluxetable}{Table}{Tables}

\newcommand{\pc}{{\,\rm pc}}
\newcommand{\kpc}{{\,\rm kpc}}
\newcommand{\Gyr}{{\,\rm Gyr}}

\newcommand{\kms}{{\,\rm km \, s^{-1}}}
\newcommand{\Msun}{{\,\rm M_\odot}}

\begin{document}

\begin{acronym}
    \acro{LOS}{line-of-sight}
    \acro{PA}{position angle}
    \acro{SN}{supernova}
    \acro{CMC}{central mass concentration}
    \acro{ISM}{interstellar medium}
    \acro{PSF}{point spread function}
    \acro{FWHM}{full width at half maximum}
    \acro{ILR}{inner Lindblad resonance}
    \acro{BIC}{Bayesian information criterion}
\end{acronym}

\title{Barflow: Modeling Gas Kinematics of Bar-Driven Flows}

\author[0009-0002-0251-9570]{Seungwon Baek}
\affiliation{Department of Physics \& Astronomy, Seoul National University, Seoul 08826, Republic of Korea}
% \email{bsw1027@snu.ac.kr}

\author[0000-0003-4625-229X]{Woong-Tae Kim}
\affiliation{Department of Physics \& Astronomy, Seoul National University, Seoul 08826, Republic of Korea}
\affiliation{SNU Astronomy Research Center, Seoul National University, Seoul 08826, Republic of Korea}
\email{unitree@snu.ac.kr}

\author[0000-0002-5857-5136]{Taehyun Kim}
\affiliation{Department of Astronomy, Yonsei University, 50 Yonsei-ro, Seodaemun-gu, Seoul 03722, Republic of Korea}
% \email{tkim.astro@gmail.com}

\author[0000-0002-7202-4373]{Dajeong Jang}
\affiliation{Korea Astronomy and Space Science Institute (KASI), 776 Daedeokdae-ro, Yuseong-gu, Daejeon 34055, Republic of Korea}
% \email{djjang@kasi.re.kr}

\author[0000-0003-0957-6201]{SungWon Kwak}
\affiliation{Leibniz-Instit\"ut f\"ur Astrophysik Potsdam (AIP), An der Sternwarte 16, 14482, Potsdam, Germany}
% \email{skwak@aip.de}

%\correspondingauthor{Woong-Tae Kim}

\begin{abstract}
We present self-consistent $N$-body/hydrodynamical simulations of isolated, Milky Way-like galaxies with stellar and gaseous disks. By varying the initial gas fraction, we find that gas delays bar formation and, in the most gas-rich model, suppresses it by exciting higher-order spiral modes and promoting early central mass growth. Once a bar forms, continued gas inflow and gravitational torques weaken the bar and inhibit its secular slowdown, keeping bars in the gas-bearing models shorter and predominantly in the fast-bar regime, while the gas-free bar evolves into the slow-bar regime. Stars and gas respond differently to the same barred potential. While the mean stellar velocity field is well described by a smooth bisymmetric pattern, the cold and warm gases follow elongated, $x_1$-like streamlines with sharp shock-induced velocity transitions, producing sawtooth-like radial and archlike tangential velocity profiles. Despite these strong noncircular motions, their azimuthally averaged tangential velocities closely follow the circular velocity over most of the disk. By contrast, the hot gas is dominated by stellar feedback and thermal pressure and is largely decoupled from the ordered bar-driven flow. To describe these bar-driven gas flows, we introduce the \textit{barflow} model, a phenomenological description of nonsinusoidal gas streaming motions with a radially varying shock-front phase. Mock line-of-sight velocity-fitting experiments show that the \textit{barflow} model recovers the intrinsic rotation curve, bar-driven velocity perturbations, and radial gas transport more accurately than the conventional bisymmetric model.
\end{abstract}

\keywords{Barred spiral galaxies (136); Spiral galaxies (1560); Disk galaxies (391); Galaxy bars (2364); Galaxy disks (589); Galaxy kinematics (602); Galaxy rotation curves (619); Galaxy structure (622)}

\section{Introduction} \label{sec:intro}

Bars are among the most common nonaxisymmetric structures in disk galaxies. Roughly two-thirds of nearby disk galaxies, including weakly barred systems, host stellar bars \citep[e.g.,][]{deVaucouleurs91,Eskridge00,Menendez07,Buta15,Diaz16}. Although the bar fraction declines toward higher redshift \citep[e.g.,][]{Abraham99,Sheth08,Melvin14}, recent deep imaging has revealed barred galaxies at $z\sim3$ and even earlier epochs \citep[e.g.,][]{Guo23,LeConte24,LeConte26,Amvrosiadis25,Geron25,Boogaard26,Wang26}. These observations indicate that stellar bars are long-lived dynamical structures that have influenced the evolution of disk galaxies over a substantial fraction of cosmic time.

The formation and secular evolution of stellar bars are governed by the interplay among stellar disks, dark matter halos, bulges, and gaseous components. A sufficiently massive and dynamically cold stellar disk is susceptible to the global bar instability, in which nonaxisymmetric perturbations grow through repeated swing amplification sustained by wave feedback \citep{Toomre64,Jang25}. The dark matter halo subsequently absorbs angular momentum from the bar and promotes its secular growth \citep[e.g.,][]{Ostriker73,Debattista00,Athanassoula02,Martinez06,Kwak17,Kwak19,Jang23,Jang24}, whereas a sufficiently massive classical bulge can delay or suppress bar formation by interrupting the feedback loop \citep[e.g.,][]{Toomre81,BT08,Kataria18,Jang23}.

Gas introduces additional complexity because, unlike the collisionless stellar and dark matter components, it is collisional and dissipative. Its low velocity dispersion reduces the effective stability of the disk and enhances the growth of nonaxisymmetric perturbations \citep[e.g.,][]{Jog84a,Jog84b,Rafikov01,Orlova02}. Numerical studies, however, have often found the opposite effect: gas-rich disks tend to form bars later and develop weaker or shorter bars than gas-poor systems \citep[e.g.,][]{Bournaud05,Villa-Vargas10,Athanassoula13,Seo19}. These apparently conflicting results suggest that gas influences bar formation and evolution through multiple competing processes, whose relative importance remains poorly understood.

The interaction between gas and bars is reciprocal. While gas can influence the formation and secular evolution of stellar bars, bars in turn drive strong noncircular gas motions through gravitational torques, transporting gas toward galactic centers \citep[e.g.,][]{Athanassoula92,Buta96,Kim12a,Kim12b,Kim12c,Kim24b}. Bar-driven gas inflows are thought to regulate central star formation \citep[e.g.,][]{Ellison11,Lin17,Seo19,Moon21,Moon22,Moon23} and promote the buildup of nuclear structures \citep[e.g.,][]{Gadotti20,Erwin24,Gadotti25,Kwak26b,Kwak26c}. These dynamical responses make gas a sensitive tracer of bar-driven secular evolution. 

Bar-driven motions have long been recognized through the characteristic S-shaped distortion of isovelocity contours \citep[e.g.,][]{Bosma78,Huntley78,Peterson78,Peterson80,Pence81,Kormendy83,Pence84,Weliachew88}. The need to interpret these velocity distortions quantitatively has become increasingly important with advances in spatially resolved observations. Integral field and millimeter interferometric surveys have now provided spatially resolved stellar and gas velocity fields for large samples of nearby disk galaxies \citep[e.g.,][]{Sanchez12,Bundy15,Gadotti19,Emsellem22,Leroy21a,Leroy21b}. Because these observations measure only the \ac{LOS} velocity, recovering the intrinsic radial and tangential motions in the disk plane remains challenging, particularly within the bar region where projection effects and strong noncircular motions are significant \citep[e.g.,][]{Salibur26}. This has motivated the development of kinematic models that infer the intrinsic in-plane velocity field from projected \ac{LOS} observations.

A widely adopted framework for this decomposition is the bisymmetric model of \citet{SS07}, in which the bar-induced radial and tangential velocities are represented by sinusoidal $m=2$ perturbations:
\begin{subequations}\label{eq:bisymmetric}
\begin{align}
V_{R}(R,\phi)
&=
-
V_{R,2}(R)
\sin[2(\phi-\phi_b)],
\label{eq:bisymmetric_radial}\\
V_{T}(R,\phi)
&=
V_{T,0}(R)
-
V_{T,2}(R)
\cos[2(\phi-\phi_b)].
\label{eq:bisymmetric_tangential}
\end{align}
\end{subequations}
Here, $V_{T,0}$ is the mean rotation velocity, $V_{R,2}$ and $V_{T,2}$ are the amplitudes of the bar-induced radial and tangential velocities, and $\phi_b$ is the bar \ac{PA} in the disk plane. The bisymmetric model  has been implemented in fitting tools such as \texttt{DiskFit} \citep{Sellwood15}, \texttt{XookSuut} \citep{LopezCoba24a}, and \texttt{Nirvana} \citep{Zanger24}, and has been widely applied to observed velocity fields of barred galaxies \citep[e.g.,][]{Hallenbeck14,Holmes15,LopezCoba22,LopezCoba24b,Hogarth24,Kim24a}.

Its applicability to gas, however, remains uncertain. The model assumes smooth sinusoidal velocity variations with a single bar-related phase angle, which provides a reasonable description of the mean stellar velocity field \citep{Baek25}. Gas, however, is collisional and dissipative, subject to shocks and radiative cooling, and often exhibits a multiphase structure \citep[e.g.,][]{Draine93,Ferriere01,Draine11}. Consequently, its response to a barred potential may deviate systematically from the sinusoidal form assumed by the bisymmetric model. Whether this assumption biases the recovery of intrinsic gas kinematics and bar-driven gas transport therefore remains an open question. 

In this paper, we address these questions using self-consistent $N$-body/hydrodynamical simulations of isolated, Milky Way-like disk galaxies with varying initial gas fractions. We first investigate how gas regulates the formation and secular evolution of stellar bars, focusing on the evolution of the bar strength, pattern speed, length, and the transition between fast and slow bars. We then examine the bar-driven kinematics of stars and multiphase gas, and demonstrate that the gaseous response differs fundamentally from the smooth bisymmetric velocity field of the stellar component. Motivated by these results, we introduce the \textit{barflow} model, a phenomenological description of nonsinusoidal gas streaming motions with a radially varying shock-front phase. Finally, using controlled mock \ac{LOS} velocity observations, we assess how accurately the \textit{barflow} model recovers intrinsic gas kinematics and bar-driven radial gas transport in comparison with the conventional bisymmetric model.

The paper is organized as follows. In \autoref{sec:model_and_method}, we describe the numerical methods and galaxy models. In \autoref{sec:evolution}, we investigate the effects of gas on bar formation and secular evolution. In \autoref{sec:kinematics}, we analyze the bar-driven stellar and multiphase gas kinematics and introduce the \textit{barflow} model. In \autoref{sec:discussion}, we evaluate the recovery of intrinsic gas kinematics from mock \ac{LOS} velocity fields and compare the performance of the \textit{barflow} and bisymmetric models. Finally, we summarize our main conclusions in \autoref{sec:summary}. 

\section{Galaxy Model and Method}\label{sec:model_and_method}

To investigate how gas affects bar formation and secular evolution and how bars drive gas kinematics, we base our models on the gas-free galaxy model of \citet{Jang23} and introduce gaseous disks of varying mass. This section describes the galaxy models and numerical methods adopted in this work.

\subsection{Galaxy Models}\label{sec:model}

We adopt model \texttt{C10} from \citet{Jang23}, whose properties resemble those of the Milky Way, as our base galaxy model. The base model consists of a dark matter halo, a stellar disk, a classical bulge, and a central supermassive black hole with mass $M_\text{BH}=4\times10^{6}\Msun$. The dark matter halo and the bulge follow the \citet{Hernquist90} profile:
\begin{equation}\label{eq:Hernquist}
    \rho(r) = \frac{M}{2\pi} \frac{a}{r(r+a)^{3}},
\end{equation}
where $r$ is the spherical radial distance, and $M$ and $a$ denote the total mass and the scale radius of each component, respectively. We set $M_h = 1.3 \times 10^{12}\Msun$ and $a_h = 30\kpc$ for the halo, while we use $M_b =  5\times10^{9}\Msun$ and $a_b = 0.4\kpc$ for the bulge. 

The stellar disk follows an exponential–sech$^2$ vertical density profile:
\begin{equation}\label{eq:expsech2}
    \rho_{s}(R,z) = \frac{M_{s}}{4\pi z_{s} R_{s}^2}\exp \left(-\frac{R}{R_{s}}\right) \text{sech}^2\left(\frac{z}{z_{s}}\right),
\end{equation}
where $R$ is the cylindrical radius, $M_{s}$ is the total stellar mass, and $R_{s}$ and $z_{s}$ are the disk scale radius and height, respectively. For all models, we fix $R_{s} = 3\kpc$ and $z_{s}=0.3\kpc$. The vertical velocity dispersion $\sigma_z$ is determined from the vertical Jeans equation for the adopted stellar density distribution and gravitational potential.  To close the radial Jeans equation, which involves the radial velocity dispersion $\sigma_R$, we set the velocity anisotropy to $f_R \equiv \sigma_R^2/\sigma_z^2 = 2.0$, following \citet{Yurin14}.

To this base model, we add a gaseous disk to construct a suite of galaxy models with differing gas content. The gaseous disk follows the same form as \cref{eq:expsech2} but with a total gas mass $M_{g}$, a disk scale length $R_g = 3\kpc$, and a radius-dependent disk scale height $z_{g}(R)$ (see \autoref{sec:method}). We vary the gas fraction $f_{g} \equiv M_{g} / (M_{s}+M_{g})$ from $0\%$ (gas free) to $20\%$, with a fixed total disk mass $M_{d}=M_{s}+M_{g} = 5 \times 10^{10} \Msun$. \cref{tab:model_parameter} summarizes the model parameters. The number following the infix \texttt{G} in the model name indicates the gas fraction.

\begin{figure}[t]
\plotone{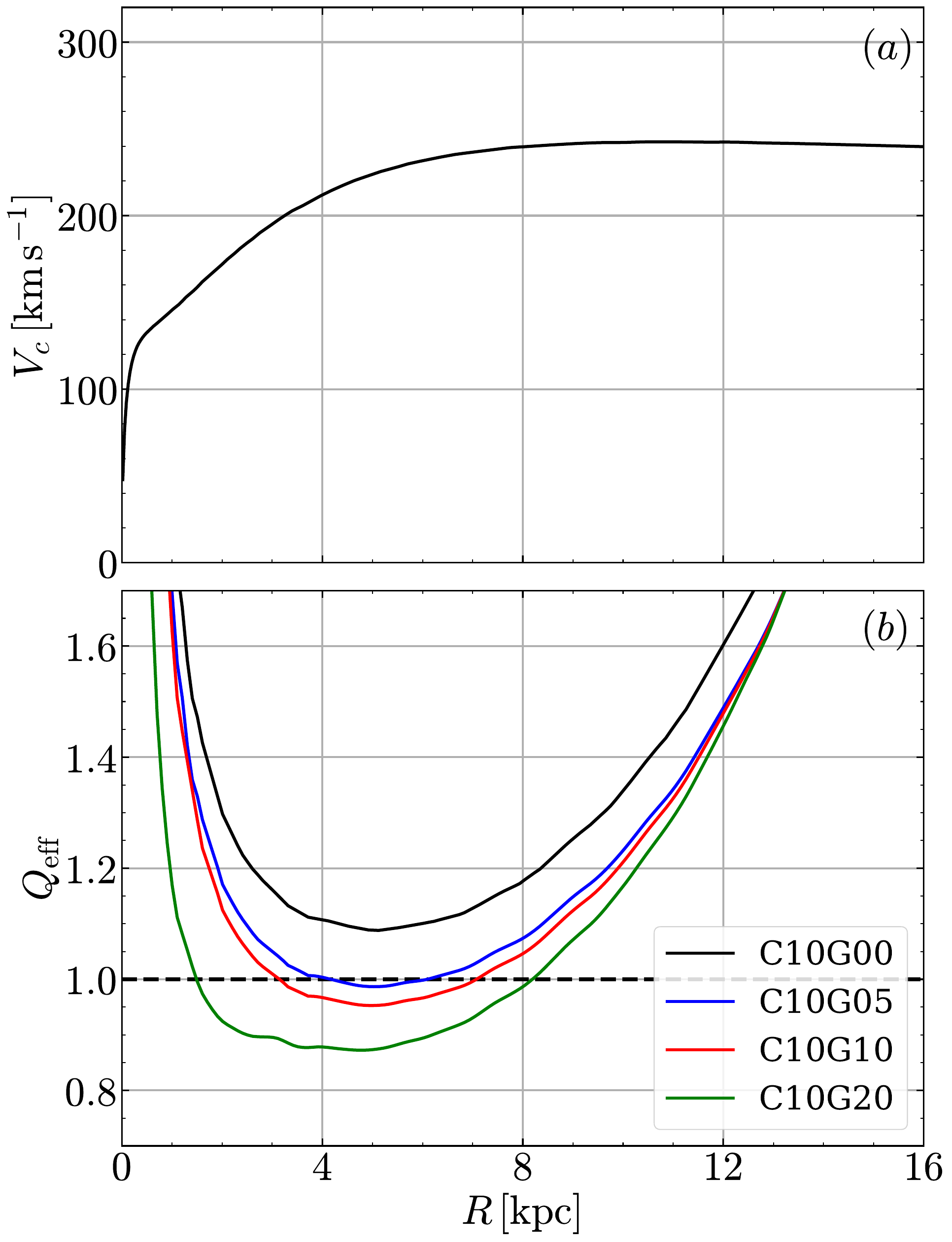}
\caption{Radial profiles of $(a)$ the circular velocity $V_c$ and $(b)$ the effective stability parameter $Q_\mathrm{eff}$ of the combined star--gas disk for models with initial gas fractions of $0\%$ (black), $5\%$ (blue), $10\%$ (red), and $20\%$ (green). The circular velocity profiles are identical for all models.}
\label{fig:vc_QT_initial}
\end{figure}

The local stability of the stellar disk, characterized by surface density $\Sigma_s$ and radial velocity dispersion $\sigma_R$, and the gaseous disk, characterized by surface density $\Sigma_g$ and sound speed $c_s$, is quantified by the \citet{Toomre64} parameters
\begin{equation}
Q_s=\frac{\kappa\sigma_R}{3.36G\Sigma_s},
\qquad
Q_g=\frac{\kappa c_s}{\pi G\Sigma_g},
\end{equation}
respectively, where $\kappa$ is the epicycle frequency.
To quantify the stability of the coupled star--gas disk, we use Equation (29) of \citet{Rafikov01} to define the effective stability parameter as 
\begin{equation}\label{eq:Qeff}
\begin{split}
Q_\text{eff}^{-1}(R)
\equiv
&\max_{q}
\left\{
\frac{2\bigl[2-e^{-q^2}I_0(q^2)\bigr]}
     {qQ_{s}(R)}
\right. \\
&\phantom{aaaaaaaaaaaaa}
\left.
+
\frac{2qs}
     {\bigl[1+(qs)^2\bigr]Q_g(R)}
\right\},
\end{split}
\end{equation}
where $q\equiv k\sigma_R/\kappa$ is the dimensionless wavenumber, $s\equiv c_s/\sigma_R$, and $I_0$ is the modified Bessel function of the first kind \citep[see also][]{Romeo11}.
\Cref{fig:vc_QT_initial} plots the initial radial profiles of the circular velocity,
$V_c=(R\,\partial\Phi/\partial R)^{1/2}$, where $\Phi$ is the gravitational potential, and the effective stability parameter $Q_\text{eff}$ for all models. Because the initial gas sound speed is lower than the stellar radial velocity dispersion, increasing the gas fraction lowers the effective disk stability and the minimum $Q_\text{eff}$. The minimum $Q_\text{eff}$ for each model is listed in the last column of \cref{tab:model_parameter}.

\subsection{Numerical Method} \label{sec:method}

The initial galaxy models are constructed using the GALIC code \citep{Yurin14}, which solves the collisionless Boltzmann equation and finds an equilibrium state for a desired density distribution by iteratively adjusting individual particle velocities. For gas-bearing models, we randomly remove stellar disk particles to reach the desired stellar mass $M_{s}=(1-f_{g})M_{d}$, and insert a self-gravitating gaseous disk with total mass $M_{g}=f_{g}M_{d}$, sound speed $c_s=10\kms$, and solar metallicity. Following \citet{Rodionov11}, we iteratively adjust the vertical scale height $z_{g}(R)$ of the gaseous disk to achieve hydrostatic equilibrium.

In all models, we fix the particle mass to $m_{h}=m_{s}=m_{b}=5\times10^{4}\Msun$ for the halo, stellar disk, and bulge, while the mass of a gas particle is set to $m_{g}=5\times10^{3}\Msun$. Accordingly, we use $N_{h}=2.6\times10^{7}$ particles for the halo and $N_{b}=10^{6}$ for the bulge in all models, whereas the numbers of stellar disk and gas particles vary between models according to the disk gas fraction $f_g$. For example, model \texttt{C10G10} consists of $N_{d}=9\times10^{5}$ stellar disk particles and $N_{g}=10^{6}$ gas particles.

We evolve our galaxy models using the GIZMO code \citep{Hopkins15}. It is a second-order-accurate hydrodynamics code based on a kernel discretization of the volume, coupled with a high-order matrix gradient estimator and a Riemann solver, which together maintain mass, energy, and momentum conservation to near machine precision. Gravity is computed with a standard tree algorithm \citep{Barnes86, Springel05} using an opening angle of $\theta=0.5$. Motivated by the effects of mass resolution and softening length on bar formation \citep{Kwak26a}, we set the gravitational softening lengths of the halo, stellar disk, and bulge particles to $50\pc$, $10\pc$, and $10\pc$, respectively. For the gas particles, we adopt fully adaptive softening lengths with a minimum value of $5\pc$.

We use a meshless finite-mass method for the evolution of gas particles, which conserves angular momentum accurately \citep{Hopkins15}. The gas particles in our models undergo radiative cooling through line emissions and heating by photoionization, cosmic-ray, and photoelectric processes \citep[e.g.,][]{Katz96, Wiersma09, Faucher-Giguere09, Hopkins18}.

We take a stochastic approach to star formation \citep[e.g.,][]{Springel03}. Star formation is allowed to occur only in sufficiently dense \citep[$n > n_{\text{crit}}=100\,\text{cm}^{-3}$;][]{Ferriere01} and self-gravitating \citep[$||\nabla \otimes {\bf v} ||^{2} + (c_s / \Delta x)^2 < 8 \pi G \rho $;][]{Semenov17, Hopkins18, Marinacci19} regions with converging flow ($\nabla \cdot {\bf v} < 0$), where $n$ is the number density, $||\nabla \otimes {\bf v} ||^{2}$ is the Frobenius norm of the velocity gradient tensor, $\rho$ is the mass density, and $\Delta x$ is the effective cell size. For a gas particle satisfying the above criteria, the probability of star formation within a time step $\Delta t$ is given by $p = 1-\exp(-\epsilon_{\text{ff}} \Delta t / t_{\text{ff}})$, where $\epsilon_{\text{ff}} = 1\%$ is the star formation efficiency \citep[e.g.,][]{Krumholz05, Krumholz07} and $t_{\text{ff}}=\sqrt{3\pi/32G\rho}$ is the free-fall timescale. The gas particle is then converted into a star particle if $p^{\ast} < p$, where $p^{\ast} \in [0,\,1)$ is a uniform random number.

\begin{deluxetable}{lcccc}
\tablecaption{
Model Parameters
\label{tab:model_parameter}
}
\tablehead{
\colhead{Name} &
\colhead{$M_s$} &
\colhead{$M_g$} &
\colhead{$f_g$} &
\colhead{$Q_{\mathrm{eff,min}}$} \\
\colhead{} &
\colhead{$[10^{10}\Msun]$} &
\colhead{$[10^{10}\Msun]$} &
\colhead{} &
\colhead{}
}
\startdata
\texttt{C10G00} &
5.00 &
0.00 &
0.00 &
1.08  \\
\texttt{C10G05} &
4.75 &
0.25 &
0.05 &
0.99 \\
\texttt{C10G10} &
4.50 &
0.50 &
0.10 &
0.95 \\
\texttt{C10G20} &
4.00 &
1.00 &
0.20 &
0.87
\enddata
\end{deluxetable}

\begin{figure*}[t]
\plotone{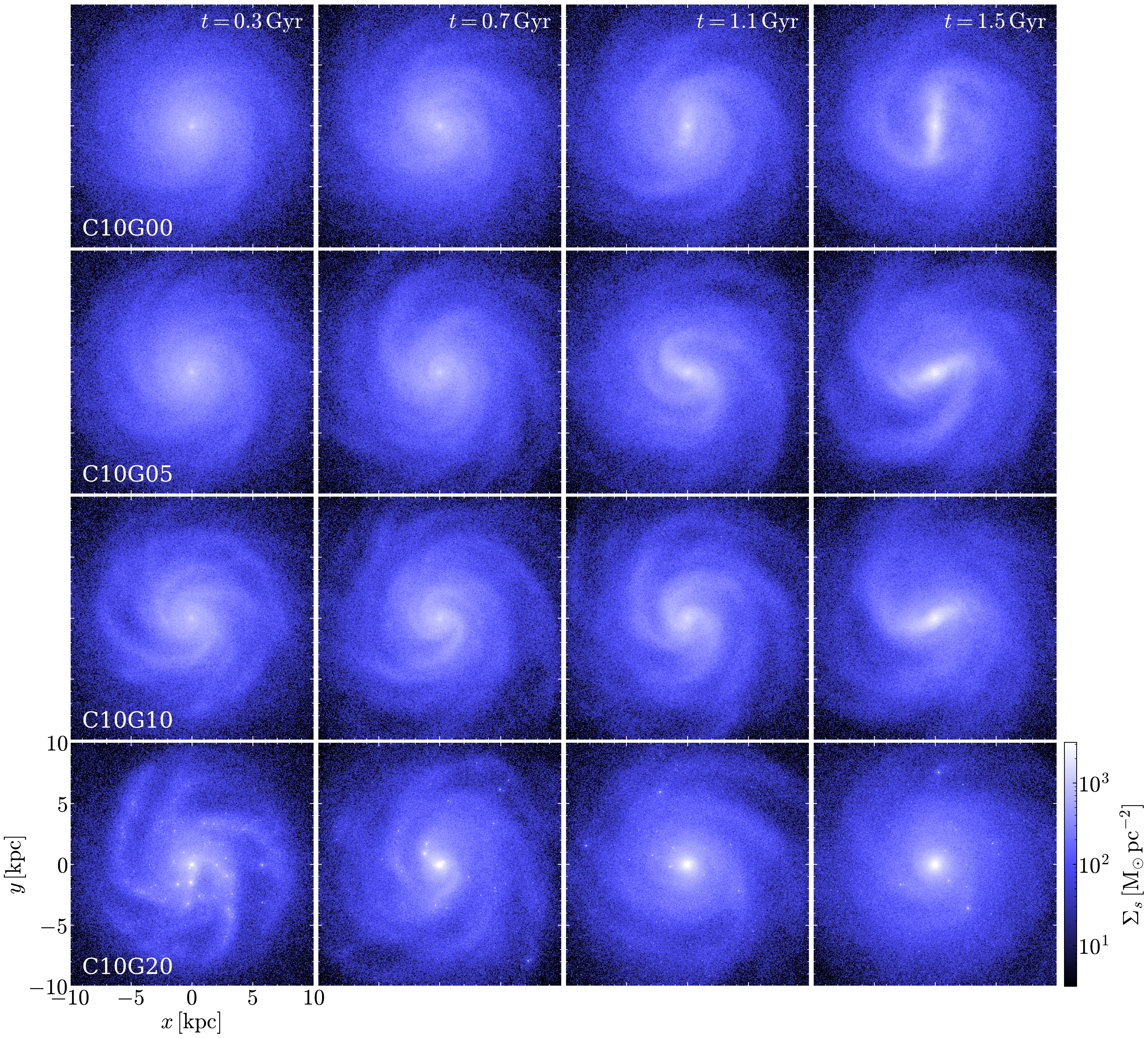}
\caption{Snapshots of the stellar surface density $\Sigma_s$ in logarithmic scale for our galaxy models at $t=0.3, \, 0.7, \, 1.1,$ and $1.5 \Gyr$ from left to right. Each row corresponds to a model with $f_{g} = 0\%, \, 5\%, \, 10\%,$ and $20\%$ from top to bottom.}
\label{fig:face_on_early}
\end{figure*}

\Ac{SN} feedback is implemented following \citet{Seo19}, using a simple prescription for momentum input and mass return to neighboring gas particles. Since each newly formed stellar particle in our models represents an unresolved star cluster, we calculate the expected number of \ac{SN} events per time step, $N_{\text{SN}}$, including both Type Ia and Type II events. The rate of Type Ia events is estimated using a delay time distribution \citep{Maoz12}, while the rate of Type II events is calculated from the \citet{Kroupa01} initial mass function together with the lifetimes of Type II progenitors \citep{Lejeune01, Seo13}. Note that, unlike Type II events, which occur only in newly formed star particles, Type Ia events can also arise from the preexisting (old) star particles that make up the initial stellar disk. For each approximately simultaneous group of \ac{SN}e, we inject mass and momentum into the neighboring gas particles within a shell radius
\begin{equation}\label{eq:r_sne}
    r_{\text{SN}} = 0.025 \, N_{\text{SN}}^{1/4} \left(\frac{n}{1\,\text{cm}^{-3}}\right)^{-1/2} \kpc.
\end{equation}
This corresponds to the shock radius at the shell-formation stage. The injected mass and momentum are distributed in proportion to the solid angle subtended by each gas particle as seen from the star particle. The total radial momentum injected is given by
\begin{equation}
    P_{\text{SN}}=2.8 \times 10^5 \, N_{\text{SN}}^{7/8} \left(\frac{n}{1\,\text{cm}^{-3}}\right)^{-0.17} \Msun\,\kms
\end{equation}
\citep[e.g.,][]{Chevalier74, Cioffi88, Seo13, Kim15}.

\section{Effects of Gas on Bar Formation and Secular Evolution}\label{sec:evolution}

In this section, we investigate the formation and evolution of stellar bars in our suite of galaxy models. We first examine how the gas fraction affects the onset time of bar formation and the resulting bar strength, and then turn to the subsequent secular evolution of the bar, characterized by its pattern speed and length.

\begin{figure}[t]
\plotone{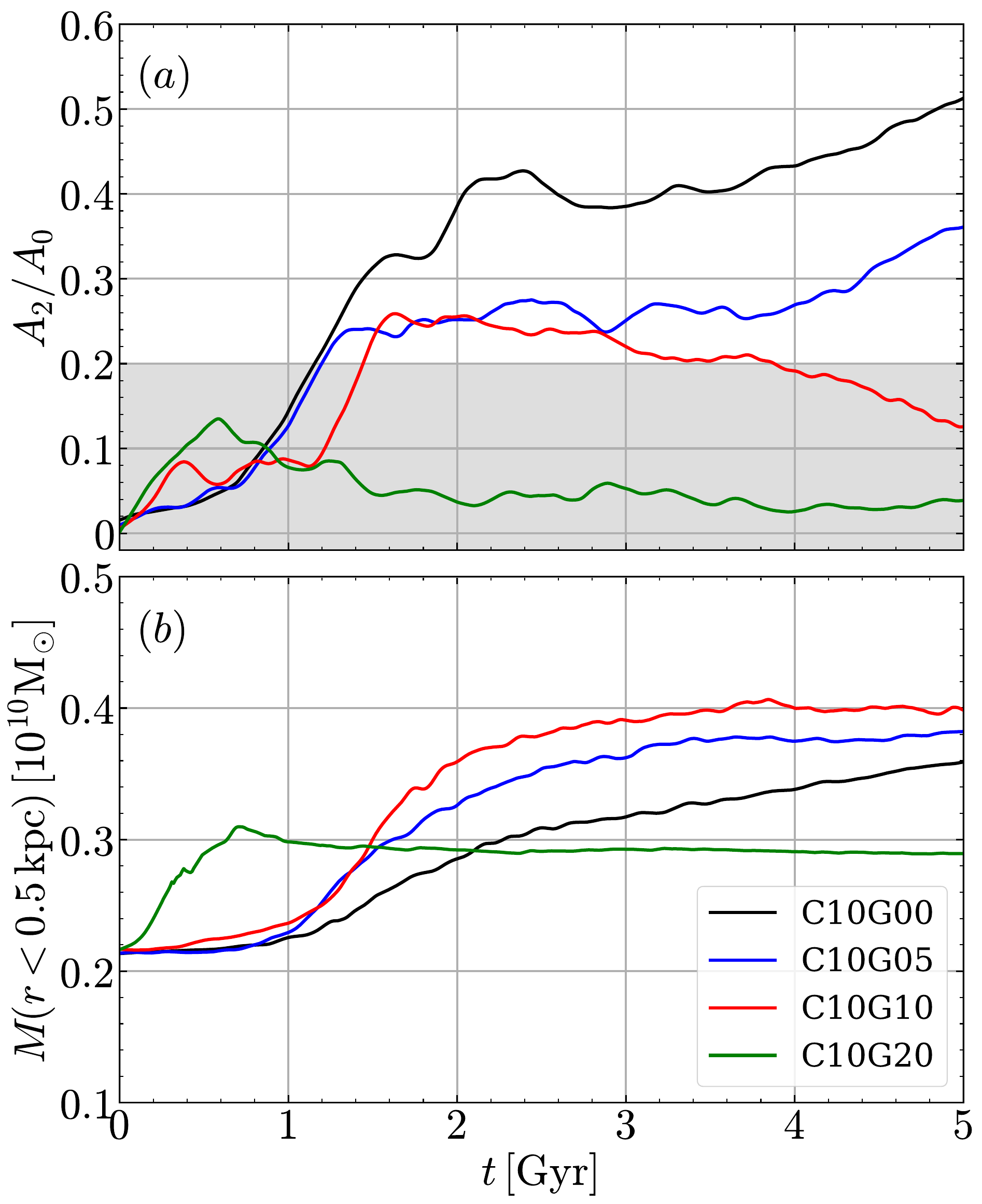}
\caption{Temporal evolution of $(a)$ the bar strength $A_2/A_0$ and $(b)$ the total baryon mass within $r<0.5\kpc$ for models \texttt{C10G00} (black), \texttt{C10G05} (blue), \texttt{C10G10} (red), and \texttt{C10G20} (green).}
\label{fig:A2A0_CMC}
\end{figure}

\subsection{Bar Formation and Strength}\label{sec:A2A0}

\cref{fig:face_on_early} plots snapshots of the stellar surface density, $\Sigma_s =\int \rho_s\,dz$, of our galaxy models at $t=0.3$, $0.7$, $1.1$, and $1.5\,\Gyr$. It is evident that models with higher gas fractions develop more prominent spiral structures at early times ($t \lesssim 1\,\Gyr$), whereas models with lower gas fractions develop distinct bar structures earlier.

To quantify the bar strength, we consider a set of concentric annuli of width $1\kpc$ centered at radius $R$, and calculate the Fourier amplitude
\begin{equation}
    a_{m}(R) = \sum_{j} \mu_{j}\exp(i m \phi_{j}), \label{eq:an}
\end{equation}
where $\phi_{j}$ and $\mu_{j}$ denote the azimuthal angle and mass of the $j$th star particle, respectively. We then define the bar strength as
\begin{equation}\label{eq:A2A0}
    \frac{A_2}{A_0} \equiv \max_R \left| \frac{a_{2}(R)}{a_{0}(R)} \right|.
\end{equation}
We classify a galaxy as barred if $A_2/A_0 \ge 0.2$; values below this threshold are attributed to ovals or spirals.

\cref{fig:A2A0_CMC}$(a)$ shows the temporal evolution of $A_2/A_0$ for our models. At early times ($t<1\Gyr$), the measured $A_2/A_0$ primarily reflects spiral structures, and models with higher gas fractions tend to exhibit larger $A_2/A_0$ with an earlier development of transient non-axisymmetric structures. However, the early enhancement of the $m=2$ mode at high $f_g$ does not necessarily lead to bar formation. While the gas-free model (\texttt{C10G00}) forms a bar at $t \simeq 1.1\Gyr$, bar formation is systematically delayed in the gas-bearing models, occurring at $t \simeq 1.2\Gyr$ and $1.4\Gyr$ for models \texttt{C10G05} and \texttt{C10G10}, respectively. In fact, the most gas-rich model \texttt{C10G20} fails to develop a bar structure at all, even though the combined disk is initially the most unstable according to $Q_\text{eff}$.

\begin{figure}[t]
\plotone{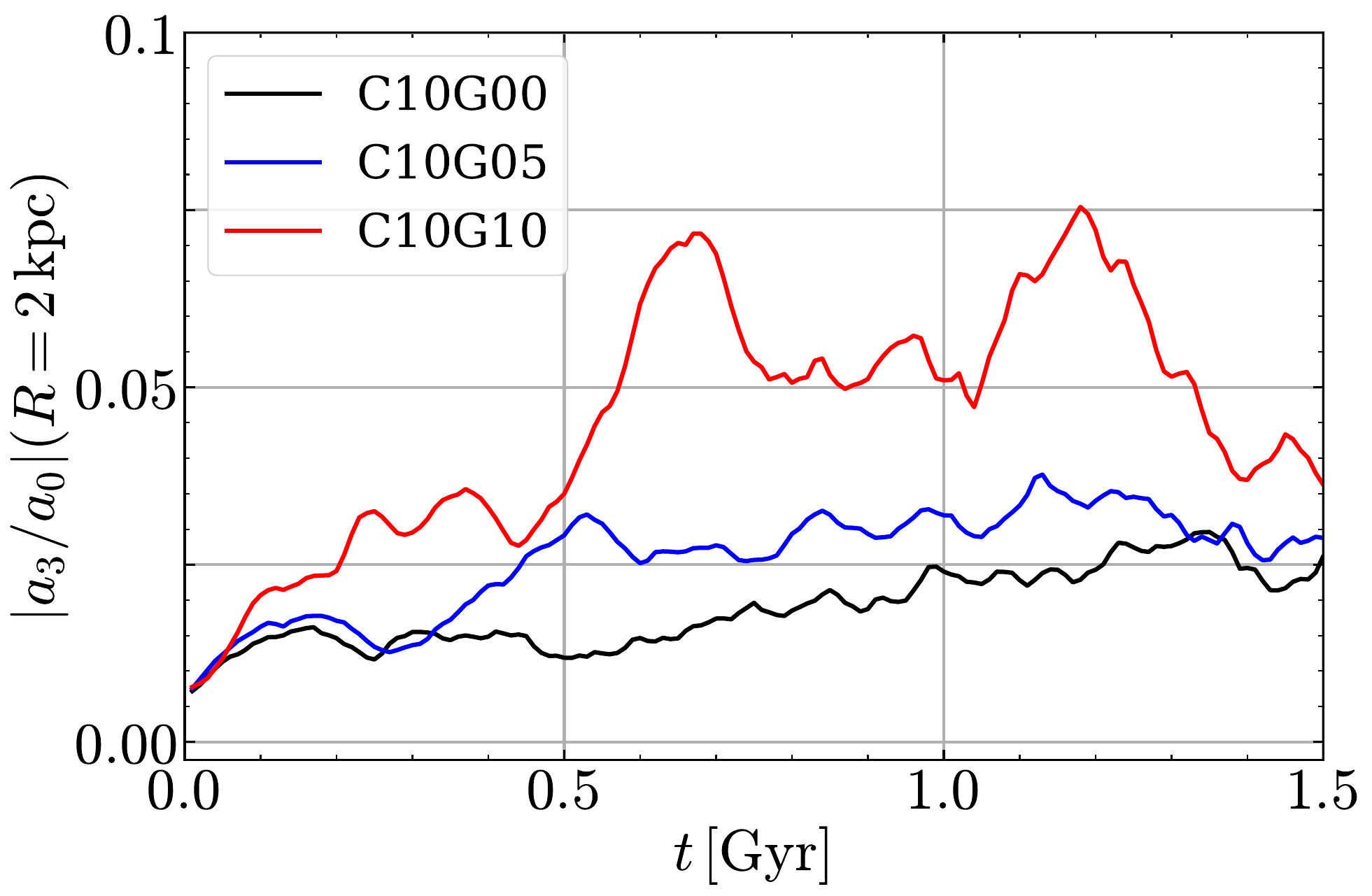}
\caption{Temporal evolution of the normalized amplitudes of the $m=3$ Fourier mode at $R=2.0\kpc$ for models \texttt{C10G00} (black), \texttt{C10G05} (blue), and \texttt{C10G10} (red).
}
\label{fig:a3a0}
\end{figure}

\begin{figure*}[t]
\plotone{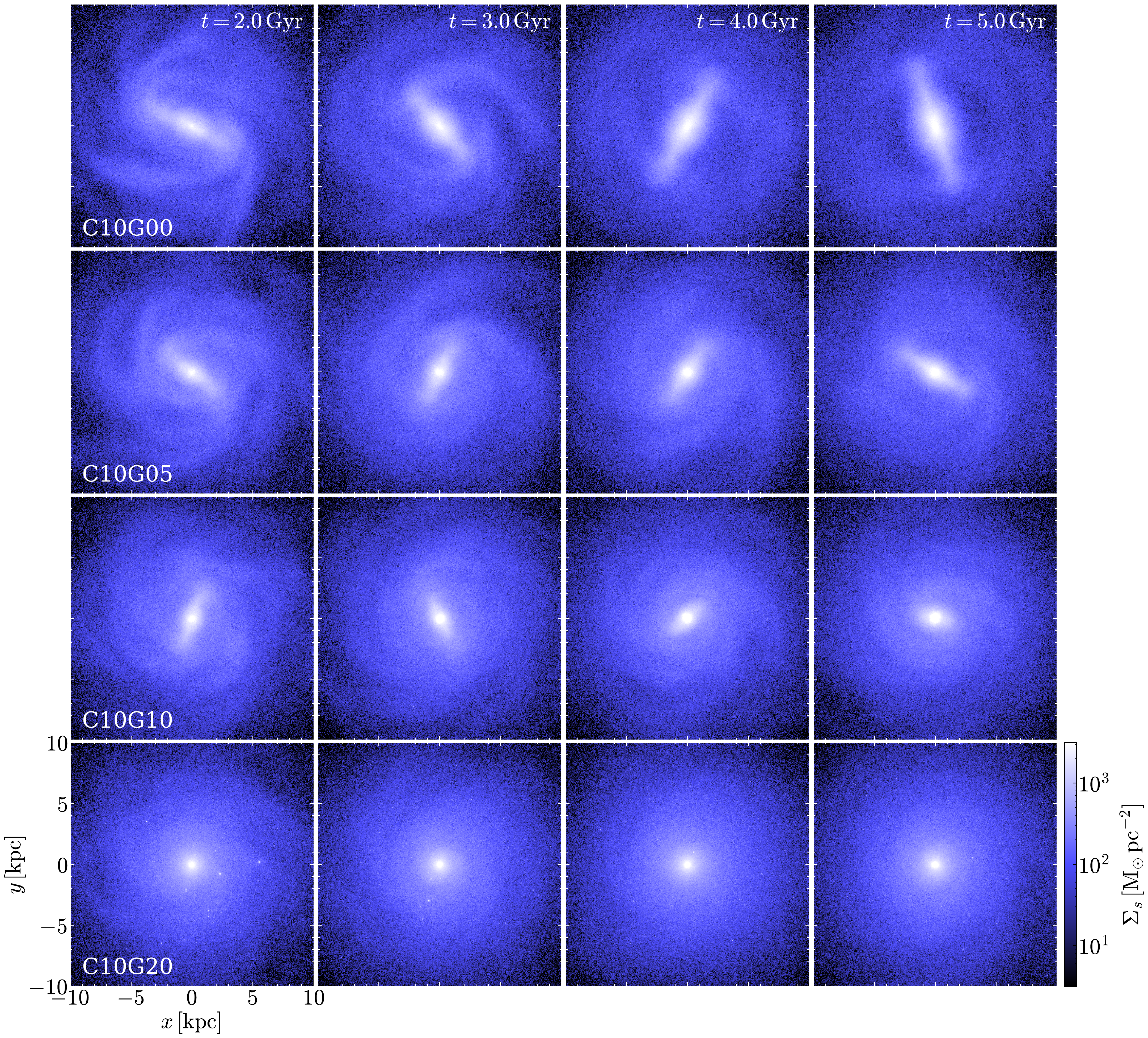}
\caption{Snapshots of the stellar surface density $\Sigma_s$ in logarithmic scale for our galaxy models at $t=2.0, \, 3.0, \, 4.0,$ and $5.0 \Gyr$ from left to right. Each row corresponds to a model with $f_{g} = 0\%, \, 5\%, \, 10\%,$ and $20\%$ from top to bottom.}
\label{fig:face_on_late}
\end{figure*}

This systematic delay of bar formation in gas-bearing models can be attributed to dynamical interference from higher-order spiral modes. \cref{fig:a3a0} plots the temporal evolution of the $m=3$ Fourier amplitude, $|a_3/a_0|$, at $R=2\kpc$ for the bar-forming models.
The presence of a dynamically cold gas component lowers the effective stability parameter $Q_{\text{eff}}$ (see \cref{fig:vc_QT_initial}$b$), making the disk not only more susceptible to gravitational perturbations but also more prone to shorter-wavelength perturbations. Consequently, disks with higher $f_g$ develop stronger $m=3$ modes (\citealt{Seo19}; see also \citealt{Kwak25}). These $m=3$ modes interact nonlinearly with the bar-forming $m=2$ mode, disrupting its coherent amplification and thereby delaying the onset of bar formation.

However, the complete suppression of bar formation in model \texttt{C10G20} requires a different dynamical explanation, rooted in the rapid gas-driven evolution of the central gravitational potential. In this extreme case, the resulting multi-armed spiral structure efficiently exerts negative torques on the large gas reservoir, driving a substantial amount of material toward the galactic center \citep[e.g.,][]{Kim14}. The early spiral-driven gas inflow rate reaches $|\dot{M}_{g}|\simeq1.0\Msun \,{\rm yr}^{-1}$ in model \texttt{C10G20}, approximately an order of magnitude larger than that in model \texttt{C10G10}.

\cref{fig:A2A0_CMC}$(b)$ shows the temporal evolution of the \ac{CMC}, defined as the total baryonic mass within $r<0.5\kpc$ \citep{Seo19}, for our galaxy models. Model \texttt{C10G20} undergoes a very steep early buildup of central mass. This rapid phase of \ac{CMC} growth greatly alters the inner orbital structure by deepening the central potential well and steeply increasing the central epicyclic frequency, thereby introducing an \ac{ILR}, where disk stars resonate with $m=2$ density waves present prior to bar formation. An \ac{ILR} acts as an absorbing boundary for inward-propagating density waves, effectively breaking the wave-reflection cycle required for repeated swing amplification \citep[e.g.,][]{Sellwood80, Toomre81, BT08}. Consequently, the feedback loop needed to trigger a global bar instability is severely inhibited, explaining why the most gas-rich model fails to form a bar at all.

The presence of gas affects not only bar formation but also its subsequent secular evolution. \cref{fig:face_on_late} plots the stellar surface density of our galaxy models at $t=2.0$, $3.0$, $4.0$, and $5.0\Gyr$. After formation, bars in the gas-bearing models are systematically weaker and shorter-lived than those in the gas-free model. For example, model \texttt{C10G05} maintains a substantially lower bar strength throughout the simulation, while the bar in model \texttt{C10G10} gradually weakens after reaching its peak and evolves into an oval distortion by $t\simeq3.9\Gyr$. This secular weakening is driven mainly by two effects of the gas. Bar-driven inflow increases the \ac{CMC}, scattering stars from the bar-supporting $x_1$ orbits and weakening the bar \citep[e.g.,][]{Hasan93, Shen04, Athanassoula05, Berentzen07}. The gas also transfers angular momentum to the bar, inhibiting its slowdown and thereby suppressing the further growth in bar length and strength \citep[e.g.,][]{Bournaud05, Beane23}.

\subsection{Bar Pattern Speed and Length}\label{sec:Omegab_Rb}

To measure the bar pattern speed, we first calculate the stellar bar \ac{PA}, $\psi(R)$, defined as
\begin{equation}\label{eq:psi}
    \psi(R) \equiv \frac{1}{2}\tan^{-1}\left[\frac{\mathrm{Im}(a_2)}{\mathrm{Re}(a_2)}\right],
\end{equation}
where $a_2$ is the complex amplitude of the $m=2$ Fourier mode (see \autoref{eq:an}). The bar pattern speed is then defined as
\begin{equation}\label{eq:Omegab}
    \Omega_b(t) \equiv \frac{d\psi(R=2.0\,\mathrm{kpc})}{dt}.
\end{equation}
We define the bar length, $R_b$, as the largest radius over which the bar \ac{PA} remains approximately constant \citep{Jang23}. The corotation radius, $R_{\rm CR}$, is determined from the condition $\Omega(R_{\rm CR})=\Omega_b$.

\begin{figure}[t]
\plotone{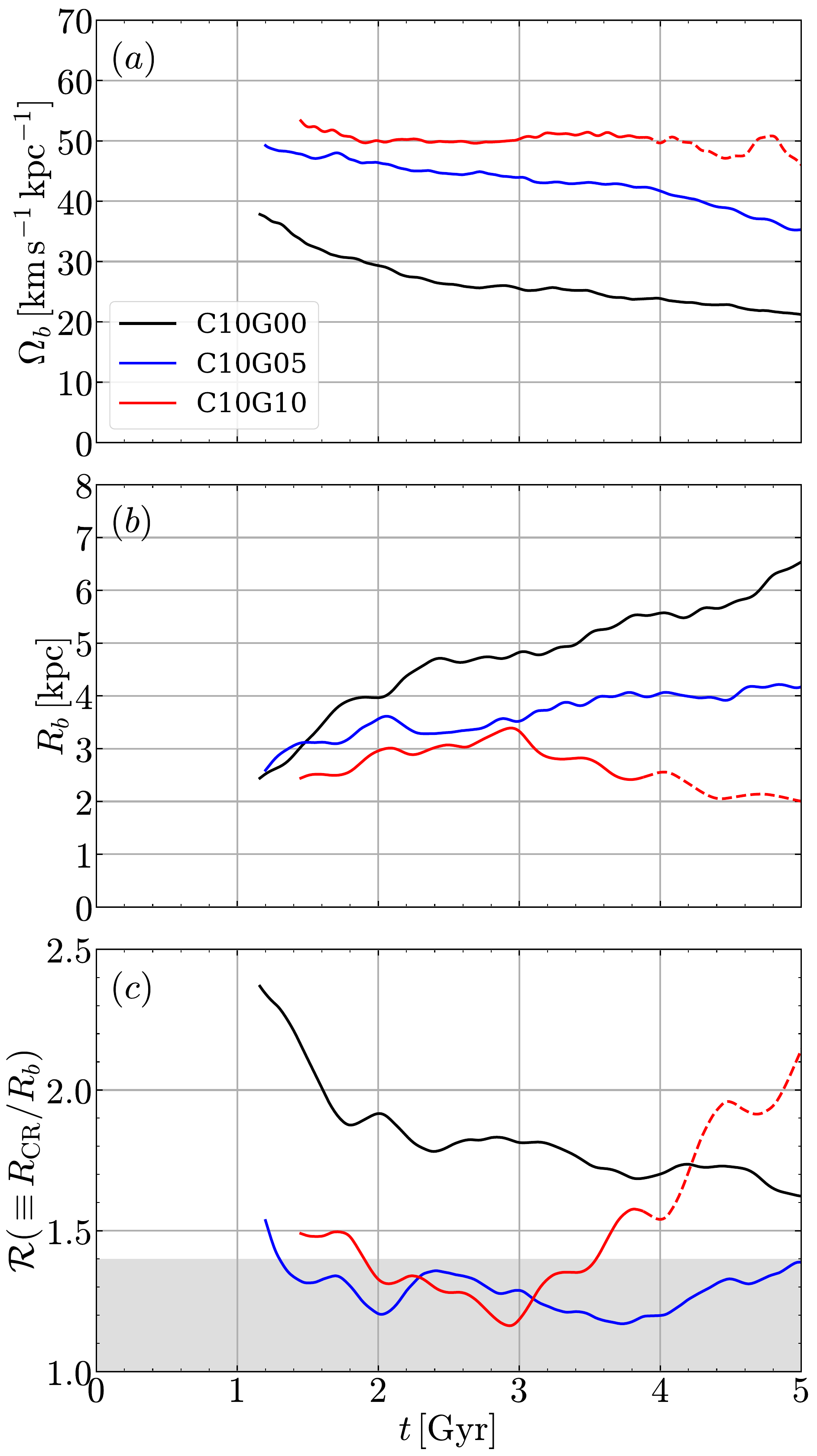}
\caption{Temporal evolution of $(a)$ the bar pattern speed, $\Omega_b$, $(b)$ the bar length, $R_b$, and $(c)$ the dimensionless ratio $\mathcal{R}=R_{\mathrm{CR}}/R_b$ for models \texttt{C10G00} (black), \texttt{C10G05} (blue), and \texttt{C10G10} (red).
For model \texttt{C10G10}, the dashed lines at $t \gtrsim 3.8 \Gyr$ indicate the period during which $A_2 / A_0 < 0.2$.
The shaded region in panel $(c)$ denotes the fast-bar regime, $\mathcal{R}\le1.4$.
}
\label{fig:Omegab_Rb_R}
\end{figure}

\Cref{fig:Omegab_Rb_R} plots the temporal evolution of $\Omega_b$, $R_b$, and the ratio $\mathcal{R} \equiv R_{\mathrm{CR}}/R_b$ for the bar-forming models. At the onset of bar formation, models with higher $f_g$ exhibit a higher pattern speed. This behavior is a direct consequence of the spiral-driven \ac{CMC} growth discussed in \autoref{sec:A2A0}. Similar to the effect of a massive classical bulge, the early accumulation of gas in the central region deepens the gravitational potential and increases the local angular and epicyclic frequencies. As a result, when the global bar instability eventually develops, it forms within a more rapidly rotating inner disk, leading to a higher initial bar pattern speed \citep[e.g.,][]{Kataria18, Kataria19, Jang23}.

Beyond the initial formation stage, the presence of gas substantially suppresses the secular slowdown of the bar. While the gas-free model \texttt{C10G00} exhibits a continuous decline in $\Omega_b$, model \texttt{C10G10} maintains an almost constant pattern speed of $\Omega_b \simeq 50\,{\rm km\,s^{-1}\,kpc^{-1}}$ throughout its lifetime. A similar trend was reported by \citet{Beane23}, who found that bars in gas-rich galaxies maintain nearly constant pattern speeds, whereas collisionless bars steadily slow down through angular momentum transfer to the dark matter halo. They attributed this behavior to gravitational torques from the gas, which counteract the halo-induced braking and effectively lock the bar pattern speed \citep[see also][]{Bournaud05,Kwak26b}.

The gas fraction required to suppress the bar slowdown differs between the two studies. \citet{Beane23} found that a gas fraction of only about $5\%$ is sufficient to maintain a nearly constant pattern speed, whereas a comparable effect appears clearly only in model \texttt{C10G10} in our simulations.
This discrepancy likely reflects differences in the galaxy models and initial gas distributions rather than a universal gas-fraction threshold. In particular, our simulations begin with an exponential gas disk, whereas \citet{Beane23} introduced a gas disk with a central hole after the bar had already formed. The threshold for suppressing bar slowdown is therefore expected to be model dependent.

Because the bar pattern speed largely determines the extent of the bar, the suppressed slowdown in the gas-bearing models also limits their bar growth, as shown in \Cref{fig:Omegab_Rb_R}$(b)$. Although all three models have similar bar lengths of $R_b \simeq 2.6$--$2.8\,\kpc$ at the onset of bar formation, their subsequent evolution depends on the gas fraction. The bar in model \texttt{C10G00} grows steadily with time, whereas its growth is weaker in model \texttt{C10G05} and is almost entirely suppressed in model \texttt{C10G10}. This behavior follows naturally from the evolution of the corotation radius, $R_{\rm CR}$ \citep[e.g.,][]{Contopoulos80, Contopoulos89}. In the gas-free model, the declining pattern speed shifts $R_{\rm CR}$ outward, providing room for the bar to grow. In contrast, the nearly constant pattern speed in the gas-bearing models keeps $R_{\rm CR}$ nearly fixed, thereby limiting the secular growth of the bar.

Bars are commonly classified as ``fast'' or ``slow'' according to whether $\mathcal{R}$ is smaller or larger than 1.4 \citep{Debattista00}. The evolution of $\mathcal{R}$ shown in \Cref{fig:Omegab_Rb_R}$(c)$ reflects the coupled evolution of the bar pattern speed and bar length. While the gas-free model evolves within the slow-bar regime \citep[see also][]{Jang23}, the gas-bearing models remain in the fast-bar regime throughout most of their evolution. Their higher pattern speeds limit the secular growth of the bar, preventing $\mathcal{R}$ from increasing substantially. These results support the idea that gas helps maintain fast bars by reducing the secular slowdown of the bar \citep{Beane23}, consistent with the predominance of fast bars in nearby disk galaxies \citep[e.g.,][]{Aguerri15, Guo19}.

\begin{figure*}[t]
\plotone{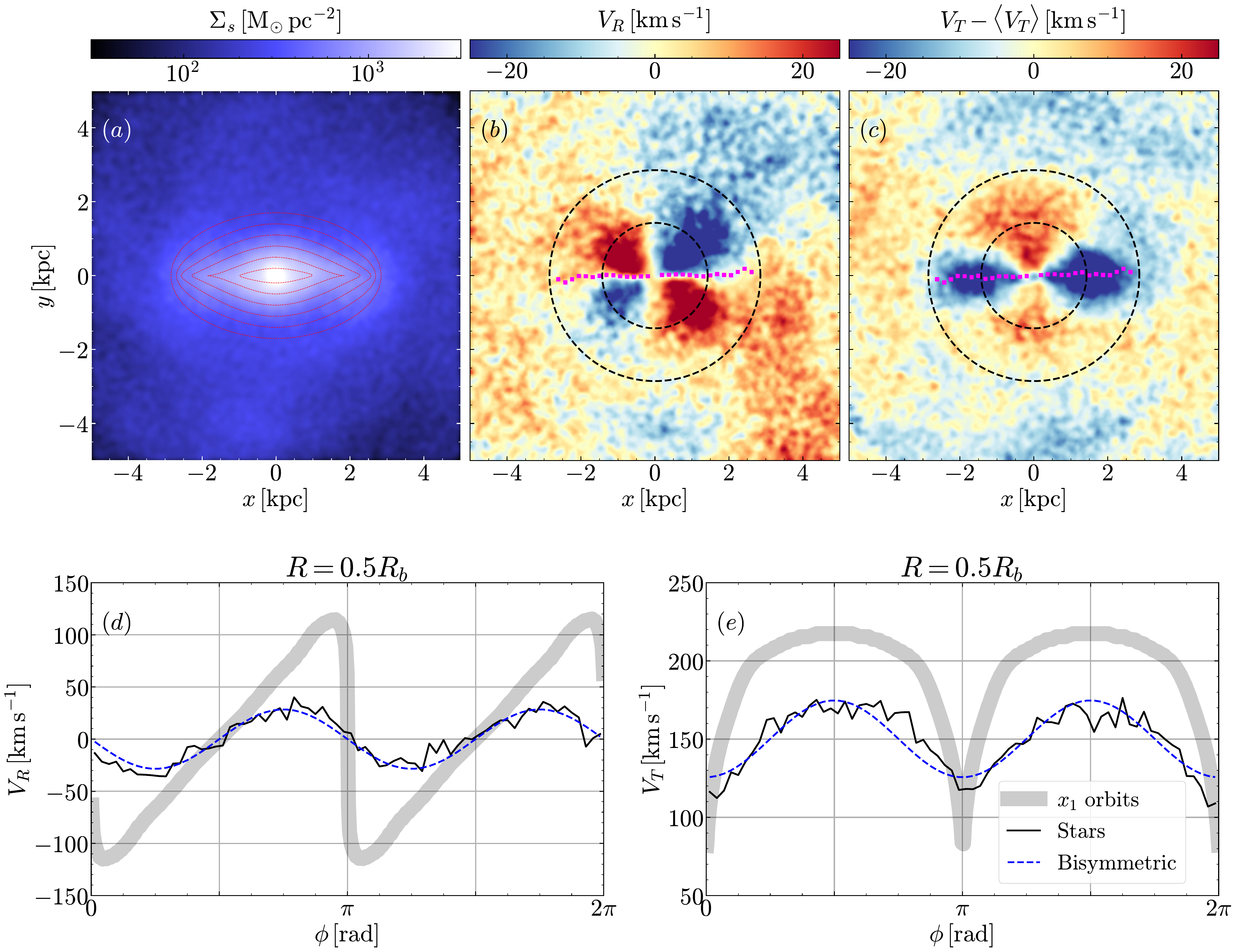}
\caption{Stellar kinematics of model \texttt{C10G10} at $t=2\Gyr$. Panels $(a)$--$(c)$ show the face-on distributions of the stellar surface density, $\Sigma_s$, the mass-weighted mean radial velocity, $V_R$, and the tangential velocity perturbation, $V_T-\langle V_T\rangle$, respectively. The bar major axis is horizontal. The red curves in panel $(a)$ show representative $x_1$ orbits. The black dashed circles in panels $(b)$ and $(c)$ mark $R=0.5R_b$ and $R=R_b$. The magenta markers indicate the stellar kinematic phase angle. Panels $(d)$ and $(e)$ show the azimuthal profiles of $V_R$ and $V_T$ measured in an annulus centered at $R=0.5R_b$ with width $\Delta R=0.1\kpc$. The gray curves correspond to the representative $x_1$ orbits, the black curves to the mass-weighted mean stellar velocities, and the blue dashed curves to the best-fit bisymmetric model.
}
\label{fig:star_kinematics}
\end{figure*}

\section{Bar-driven Stellar and Gas Kinematics}\label{sec:kinematics}

In this section, we analyze the stellar and gas kinematics of barred galaxies formed in our simulations. We show that the cold and warm gases exhibit nonsinusoidal, $x_1$-like streaming motions, in contrast to the smoother bisymmetric velocity field of the stars. We then introduce the \textit{barflow} model, designed to capture these characteristic bar-driven gas flows, which are intrinsically distinct from the stellar kinematics.

\subsection{Stellar Response to the Bar}\label{sec:star_kinematics}

\Cref{fig:star_kinematics} plots the face-on stellar surface density of model \texttt{C10G10} at $t=2\Gyr$, overlaid with representative $x_1$ orbits, two-dimensional maps of velocity perturbations, and azimuthal velocity profiles at $R=0.5R_b$.
In the disk plane, the bar induces coherent quadrupole patterns in both the radial velocity $V_R$ and the tangential velocity perturbation $V_T-\langle V_T\rangle$, where $V_R \equiv \int v_R\rho\,dz/\int\rho\,dz$ and $V_T \equiv \int v_T\rho\,dz/\int\rho\,dz$ are the vertically averaged radial and tangential velocities, respectively, and the angle brackets $\langle\,\rangle$ denote the azimuthal average at each radius. Such quadrupole patterns are commonly found in both simulations \citep[e.g.,][]{Bovy19,Duran24,Ghosh25,Duran26} and observations  \citep[e.g.,][]{Queiroz21,Gaia23}.
We define the radial and tangential kinematic phase angles as the azimuths where $V_R$ changes sign from positive to negative ($V_R=0$) and where $V_T-\langle V_T\rangle$ reaches its minimum ($V_{T,\rm min}$), respectively. Note that these kinematic phase angles remain nearly independent of radius and closely follow the bar major axis, as indicated by the magenta markers. 

The $x_1$ orbit family forms the primary orbital backbone of stellar bars and consists of elongated orbits aligned with the bar major axis \citep{BT08}. Owing to their large eccentricities, individual $x_1$ orbits exhibit strong noncircular motions. At a fixed radius, they produce sawtooth-like radial velocity profiles and archlike tangential velocity profiles as functions of azimuth, as illustrated by the thick gray curves in \Cref{fig:star_kinematics}$(d)$ and $(e)$. In contrast, the mean stellar velocity profiles measured at $R=0.5R_b$ (thin black curves) are much smoother. They are well described by sinusoidal functions (blue dashed curves), together with the nearly radius-independent kinematic phase angle. This smooth, bisymmetric velocity field underlies the success of the bisymmetric model of \citet{SS07} in describing the large-scale stellar kinematics of barred galaxies.

The smooth mean velocity field arises from collisionless orbital mixing. The stellar orbits in the bar are not limited to the $x_1$ family; box orbits and other resonant orbit families also contribute substantially \citep[e.g.,][]{Voglis07,Valluri16,Gajda16}. Averaging over this orbital distribution smooths the sharp velocity variations of individual $x_1$ orbits, producing the nearly sinusoidal profiles and coherent quadrupole patterns seen in the mean stellar kinematics \citep{Baek25}.

\begin{figure}[t]
\plotone{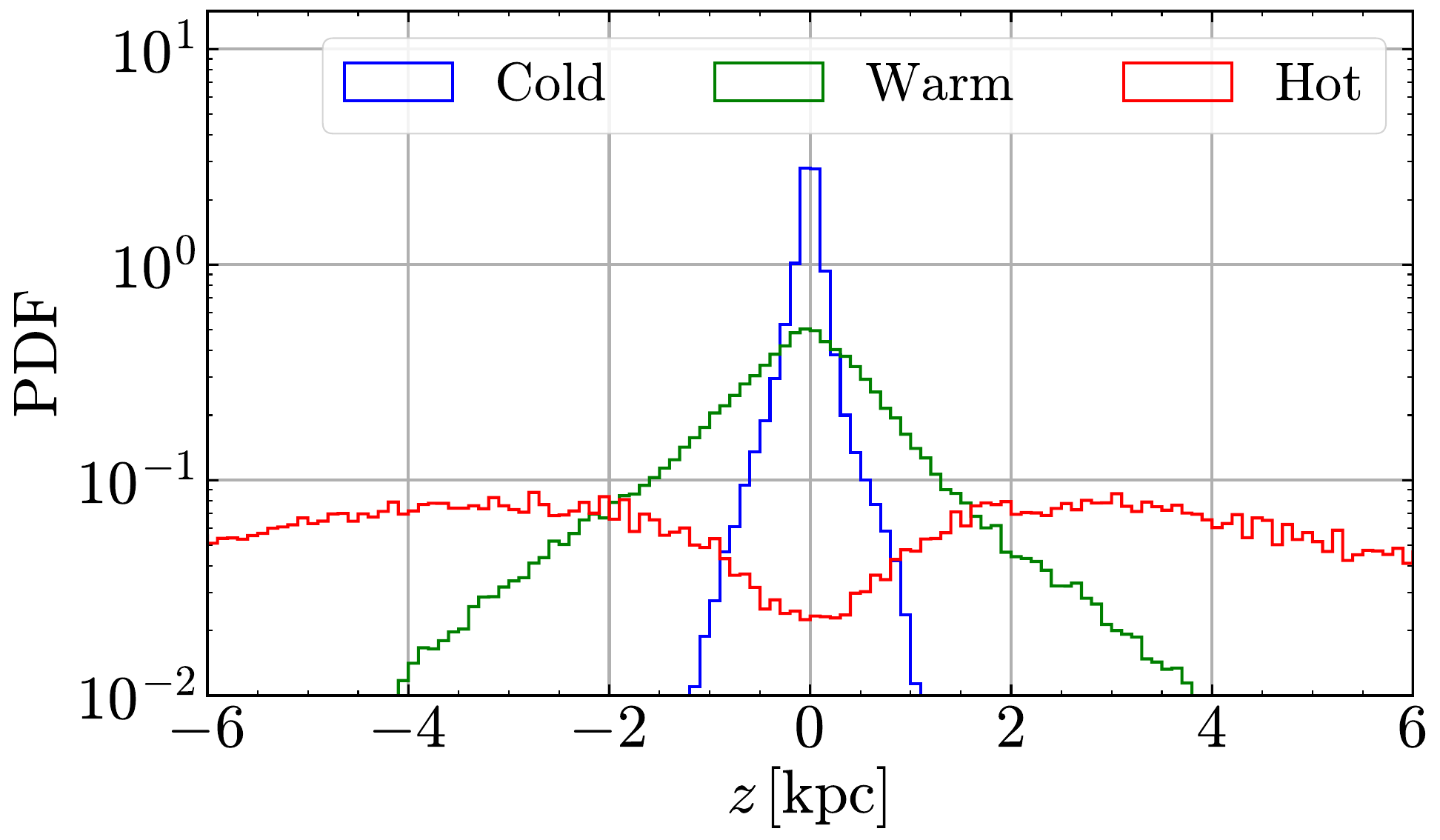}
\caption{Probability density functions of the vertical position, $z$, for cold ($T<10^{3}\,\mathrm{K}$; blue), warm ($10^{3}\,\mathrm{K}\le T<10^{4.5}\,\mathrm{K}$; green), and hot ($T\ge10^{4.5}\,\mathrm{K}$; red) gas in model \texttt{C10G10} at $t=2\,\Gyr$. Each distribution is normalized to the unit area.
}
\label{fig:gas_z_distribution}
\end{figure}

\begin{figure*}[t]
\centering
\includegraphics[width=0.8\textwidth]{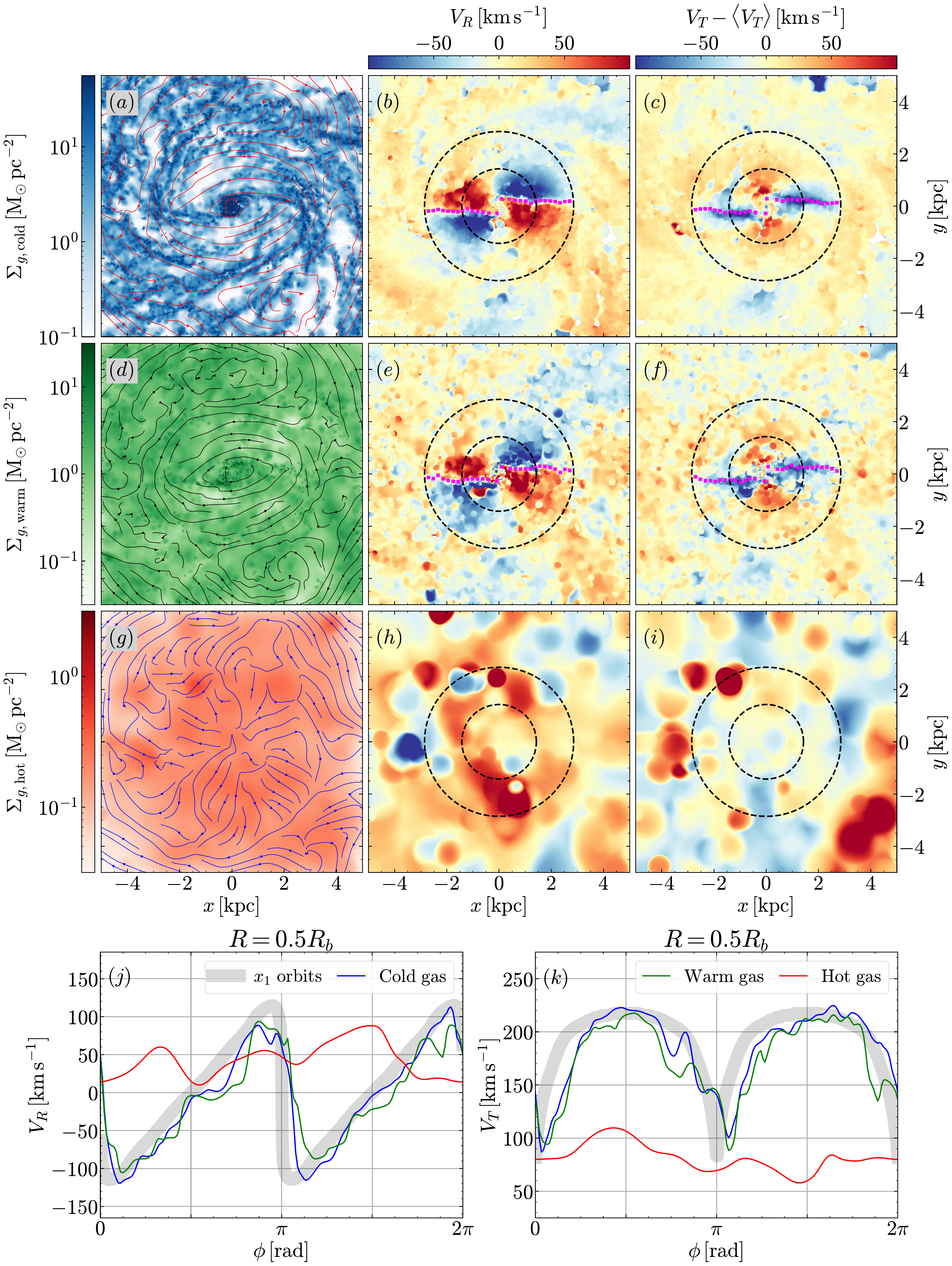}
\caption{Multiphase gas kinematics of model \texttt{C10G10} at $t=2\Gyr$. The upper three rows show the face-on distributions of the gas surface density, $\Sigma_g$, the density-weighted mean radial velocity, $V_R$, and the tangential velocity perturbation, $V_T-\langle V_T\rangle$, from left to right. The rows correspond to the cold ($T<10^3\,\mathrm{K}$), warm ($10^3\,\mathrm{K}\le T<10^{4.5}\,\mathrm{K}$), and hot ($T\ge10^{4.5}\,\mathrm{K}$) gas, respectively. The bar major axis is horizontal. Streamlines overlaid on the surface-density maps show the instantaneous gas flow in the frame corotating with the bar. The black dashed circles mark $R=0.5R_b$ and $R=R_b$. The magenta markers indicate the gas kinematic phase angle. The bottom row shows the azimuthal profiles of the density-weighted mean $V_R$ and $V_T$ measured in an annulus centered at $R=0.5R_b$. The gray curves correspond to representative $x_1$ orbits, while the blue, green, and red curves correspond to the cold, warm, and hot gas, respectively.}
\label{fig:gas_kinematics}
\end{figure*}

\begin{figure}[t]
\plotone{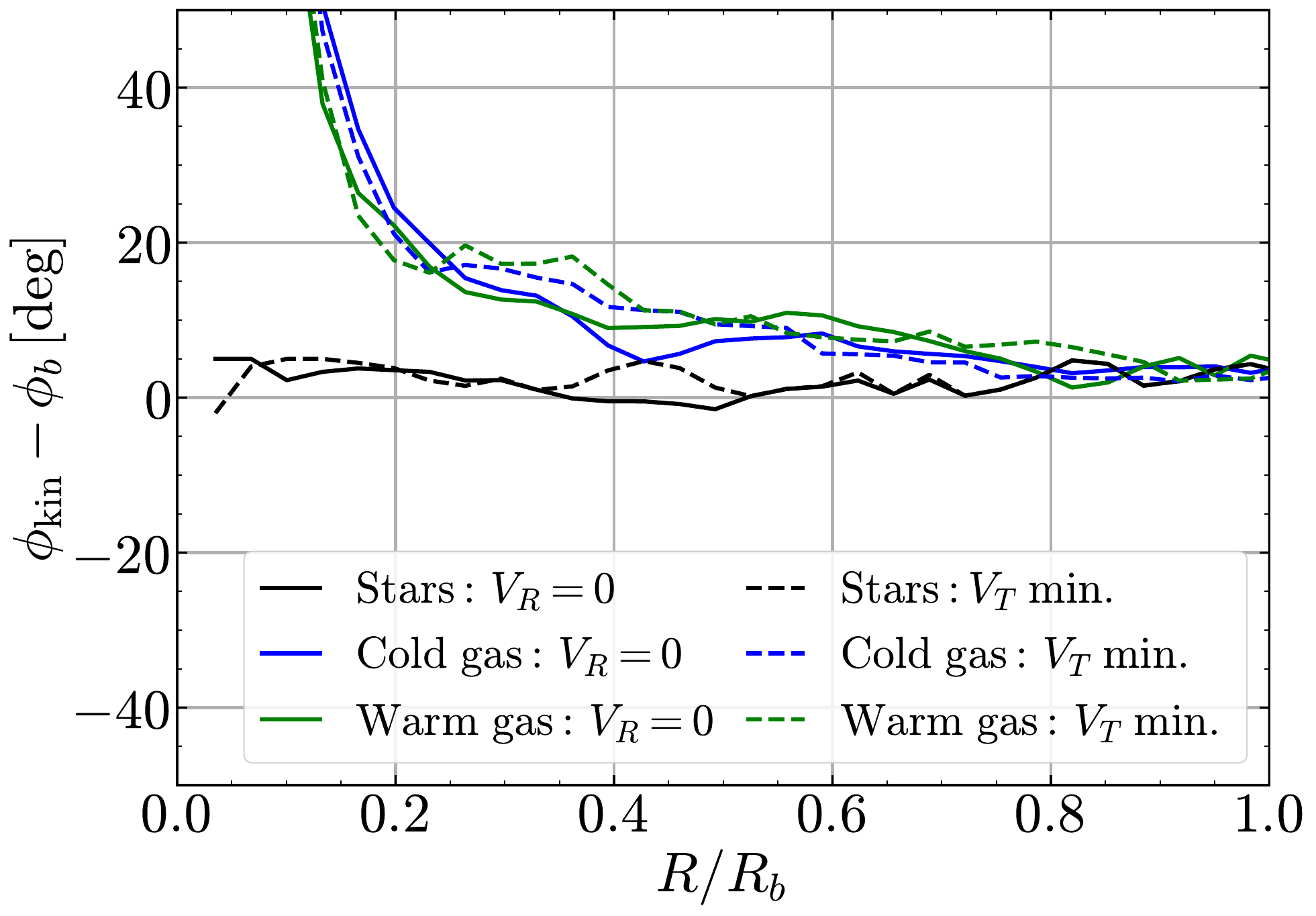}
\caption{Radial variation of the kinematic phase angle, $\phi_{\rm kin}$, relative to the stellar bar phase angle, $\phi_b$, in model \texttt{C10G10} at $t=2\Gyr$. Positive values of $\phi_{\rm kin}-\phi_b$ correspond to the leading side of the bar.
}
\label{fig:kinematic_angle}
\end{figure}

\subsection{Multiphase Gas Response to the Bar}\label{sec:gas_kinematics}

The gaseous component responds to the bar differently from the stellar component because it is collisional and dissipative and is subject to stellar feedback. \ac{SN} explosions and stellar winds continuously inject thermal and turbulent energy into the \ac{ISM}, while radiative cooling removes it, naturally producing a multiphase medium. Following \citet{Tress20}, we divide the gas into three thermal phases: cold ($T<10^3\,\mathrm{K}$), warm ($10^3\,\mathrm{K}\le T<10^{4.5}\,\mathrm{K}$), and hot ($T\ge10^{4.5}\,\mathrm{K}$).  At $t=2\Gyr$ in model \texttt{C10G10}, the cold, warm, and hot phases contain 57\%, 31\%, and 12\% of the total gas mass, respectively. As shown in \Cref{fig:gas_z_distribution}, the cold gas forms a geometrically thin disk, the warm gas has a broader vertical distribution, and the hot gas extends far above the midplane owing to stellar feedback and its high thermal pressure.

\Cref{fig:gas_kinematics} shows the face-on distributions of the gas surface density, radial velocity, and tangential velocity perturbation for each phase. All velocities are density-weighted vertical averages.
The figure also shows the azimuthal velocity profiles measured at $R=0.5R_b$. The gas response depends strongly on its thermal phase. The cold and warm phases exhibit coherent bar-driven streaming motions and follow elongated, $x_1$-like streamlines in the frame corotating with the bar \citep[e.g.,][]{Athanassoula92,Kim12a,Kim12b,Sormani15,Sormani18,Tress20}. The associated gas inflow forms a nuclear ring or disk composed primarily of cold gas, within which the gas streamlines become $x_2$-like, with their major axes approximately perpendicular to the bar major axis. \citep[e.g.,][]{Athanassoula92,Wada94,Kim12a,Kim12b}. By contrast, the hot gas is dominated by feedback- and pressure-driven motions. Compared to the stellar component, however, the cold and warm gases display much sharper velocity variations. While the stellar velocity field is characterized by smooth quadrupole patterns with nearly sinusoidal azimuthal variations (\Cref{fig:star_kinematics}), the gas exhibits abrupt velocity jumps associated with shocks along the leading edges of the bar.

Another major difference from the stellar response is the radial variation of the kinematic phase angle of the gas. The stellar phase angle remains nearly constant with radius, whereas the gas phase angle varies systematically across the bar region. 
The magenta markers in \Cref{fig:gas_kinematics}$(b,c,e,f)$ trace the gas kinematic phase angle and closely follow the leading dust lanes. This behavior is quantified in \Cref{fig:kinematic_angle}, where the cold and warm phases show large offsets from the bar major axis in the inner disk that gradually decrease toward the bar end, reflecting the curvature of the leading shocks \citep[e.g.,][]{Athanassoula92,Wada94,Kim12a,Kim12b}. The cold and warm phases closely follow each other, indicating that they trace the same shock structure. Moreover, the radial velocity reversal nearly coincides with the tangential velocity minimum (\cref{fig:gas_kinematics}$j,k$), suggesting that both arise from the same shock front.

These properties are also evident in the azimuthal velocity profiles. The cold and warm gases exhibit similar $x_1$-like kinematic patterns, characterized by sawtooth-like radial velocity profiles and archlike tangential velocity profiles, although they differ in detail. Pressure gradients and shocks modify the $x_1$-like kinematics, while dissipation maintains sharp velocity transitions. As a result, the gas velocity profiles deviate substantially from the sinusoidal forms expected from a simple bisymmetric model.

The warm gas also exhibits substantial small-scale velocity fluctuations driven by turbulence and stellar feedback \citep[e.g.,][]{Walch15,Padoan16,Kim17}. Nevertheless, its large-scale streaming pattern remains similar to that of the cold gas because both phases are predominantly confined to the disk and respond to the same bar-induced gravitational forcing \citep[e.g.,][]{Kim12a,Kim12b,Sormani15}. Turbulence and feedback mainly introduce local, incoherent velocity fluctuations in the warm phase.
The hot phase behaves quite differently. Because it is primarily supported by thermal pressure and stellar feedback, much of the hot gas resides above the disk (\Cref{fig:gas_z_distribution}) and is largely decoupled from the ordered bar-driven flow. Instead, its velocity field is dominated by localized turbulent and pressure-driven motions, with a weak large-scale radial outflow \citep[e.g.,][]{Li17,KimCG18,Fielding18}.

To summarize, the multiphase comparison highlights three key properties that any phenomenological model of bar-driven gas kinematics should capture. First, the cold and warm gases exhibit intrinsically nonsinusoidal velocity profiles, with the characteristic sawtooth-like radial motions and archlike tangential motions. Second, the kinematic phase angle varies systematically with radius as the leading shocks curve across the bar. Third, the model should focus on the cold and warm phases, which together account for $\sim90\%$ of the gas mass and trace the large-scale gravitational response to the bar, rather than the hot phase, whose kinematics are dominated by stellar feedback.

\begin{figure}[t]
\plotone{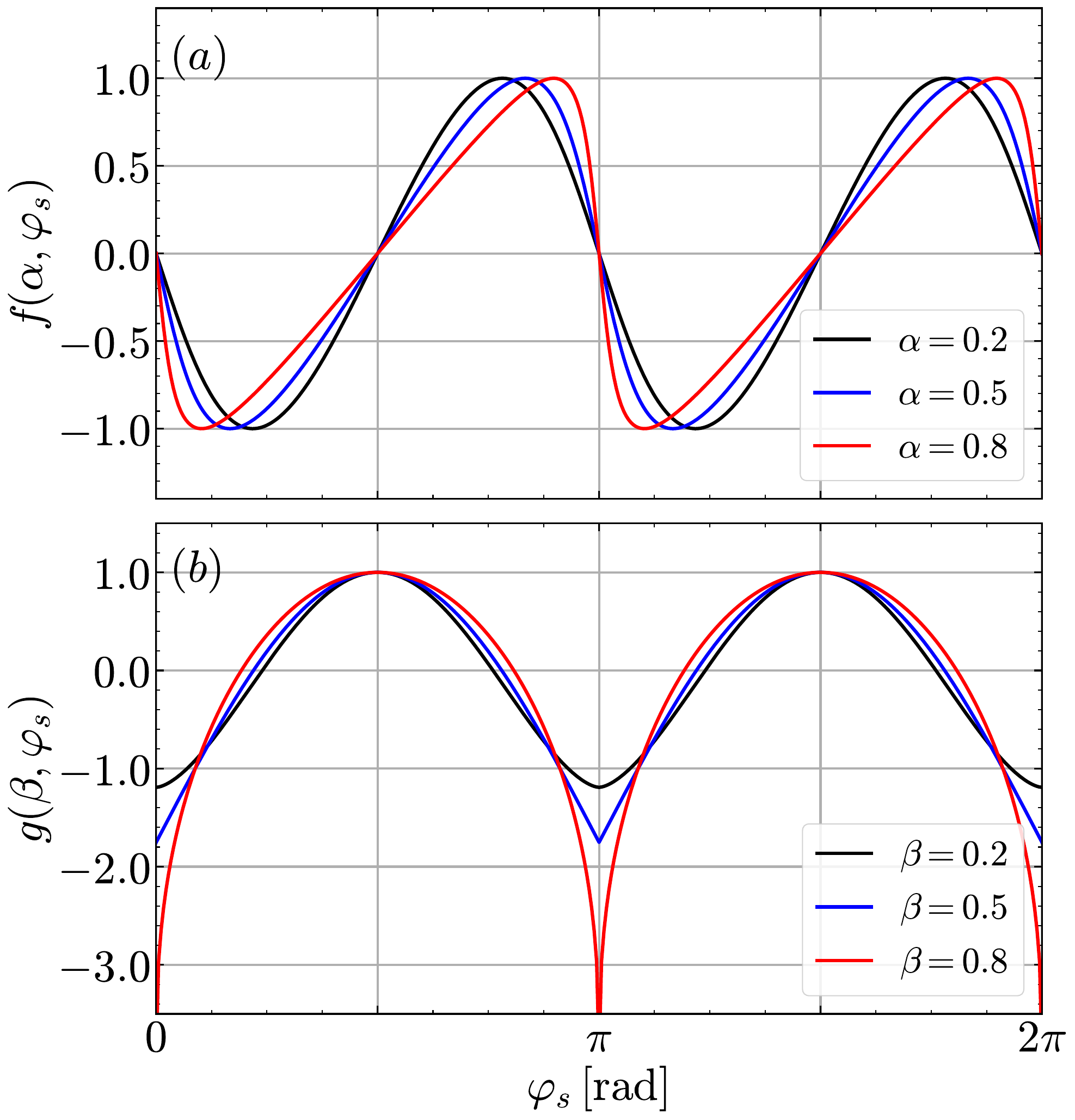}
\caption{Azimuthal dependence of the dimensionless \textit{barflow} functions. ($a$) Radial function $f$ for $\alpha=0.2$, $0.5$, and $0.8$. ($b$) Tangential function $g$ for $\beta=0.2$, $0.5$, and $0.8$. Larger values of $\alpha$ and $\beta$ produce progressively sharper sawtooth-like and archlike profiles, respectively.}
\label{fig:analytic_func}
\end{figure}

\subsection{The Barflow Model for Bar-driven Gas Flows}\label{sec:barflow_model}

As discussed in \autoref{sec:gas_kinematics}, the cold and warm gases exhibit bar-driven velocity fields that differ qualitatively from the stellar response. Their radial and tangential velocity profiles are intrinsically nonsinusoidal, showing sawtooth-like and archlike angular variations, respectively. In addition, the kinematic phase angle varies systematically with radius as the leading shocks curve across the bar. These features cannot be reproduced by the conventional bisymmetric model (\autoref{eq:bisymmetric}). We therefore introduce a phenomenological model for bar-driven gas flows, which we refer to as the \textit{barflow} model.

The \textit{barflow} model parameterizes the gas velocity field in terms of the mean rotation, the amplitudes of the radial and tangential velocity perturbations, and a radially varying shock-front phase:
\begin{subequations}\label{eq:barflow}
\begin{align}
V_{R}(R,\phi) &=
\Delta V_{R}(R)\,
f\!\left[\alpha(R),\varphi_s(R)\right],
\label{eq:gas_radial_vel} \\
V_{T}(R,\phi) &=
\langle V_{T}\rangle(R)
+\Delta V_{T}(R)\,
g\!\left[\beta(R),\varphi_s(R)\right],
\label{eq:gas_tangential_vel}
\end{align}
\end{subequations}
where $\varphi_s(R)\equiv\phi-\phi_s(R)$ is the azimuth measured relative to the local shock front, with its phase angle $\phi_s(R)$ allowed to vary with radius.\footnote{Within the nuclear gas disk (or ring), $\phi_s(R)$ represents the kinematic phase of the $x_2$-like flow and does not necessarily correspond to a shock front.}
Here, $\Delta V_R(R)$ and $\Delta V_T(R)$ denote the amplitudes of the radial and tangential velocity perturbations, respectively.

We adopt the following dimensionless functions for the radial and tangential velocity perturbations:
\begin{subequations}
\begin{align}
f(\alpha,\varphi_s)
&=
-\frac{1}{\arcsin\alpha}
\sum_{n=1}^{\infty}
\frac{\alpha^n}{n}\sin(2n\varphi_s)
\nonumber\\
&=
-\frac{1}{\arcsin\alpha}
\tan^{-1}
\left[
\frac{\alpha\sin(2\varphi_s)}
     {1-\alpha\cos(2\varphi_s)}
\right],
\label{eq:radial_func}
\\
g(\beta,\varphi_s)
&=
\frac{|\sin^2\varphi_s|^{1-\beta}-C(\beta)}
     {1-C(\beta)}
\nonumber\\
&=
\frac{|\sin^2\varphi_s|^{1-\beta}
-\Gamma(1.5-\beta)/[\sqrt{\pi}\Gamma(2-\beta)]}
{1-\Gamma(1.5-\beta)/[\sqrt{\pi}\Gamma(2-\beta)]},
\label{eq:tangential_func}
\end{align}
\end{subequations}
where $C(\beta)\equiv (2\pi)^{-1} \int
|\sin\phi|^{2(1-\beta)} \, d\phi$.
The shape parameters satisfy $0<\alpha,\beta<1$. The function $f$ is normalized to have extrema of $\pm1$, while $g$ has zero azimuthal mean and a maximum value of unity. \Cref{fig:analytic_func} illustrates the dependence of $f$ and $g$ on the shape parameters. Increasing $\alpha$ and $\beta$ produces progressively sharper sawtooth-like and archlike profiles, respectively. These parameters may reflect both the underlying $x_1$-like orbital structure and the gas response to it, including shocks and pressure effects. Their values may therefore depend collectively on the bar properties and gas properties such as density, flow velocity, sound speed, turbulent velocity dispersion, and, when present, magnetic fields \citep[e.g.,][]{Athanassoula92,Kim12b,Sormani15}.

\begin{figure}[t]
\plotone{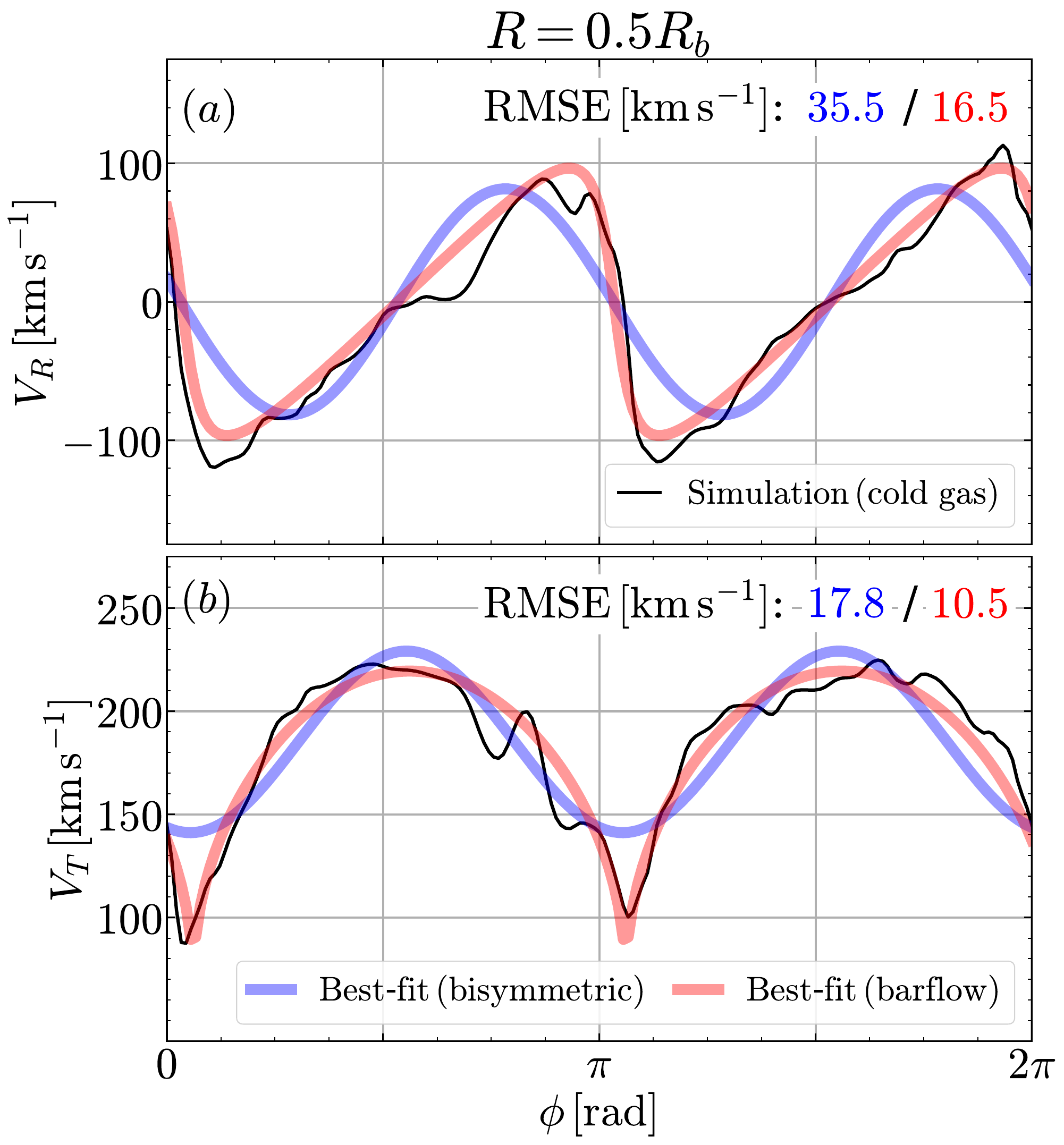}
\caption{Comparison of the bisymmetric and \textit{barflow} models for the azimuthal profiles of $(a)$ the radial velocity, $V_R$, and $(b)$ the tangential velocity, $V_T$. The black curves show the simulated cold-gas velocities in model \texttt{C10G10} at $R=0.5R_b$ and $t=2\Gyr$, while the blue and red curves show the best-fit bisymmetric and \textit{barflow} models, respectively. The best-fit \textit{barflow} parameters are $\alpha=0.79$ and $\beta=0.77$. The root-mean-square error of each fit is indicated in the corresponding panel.}
\label{fig:model_comparison_1D}
\end{figure}

\begin{figure}[t]
\plotone{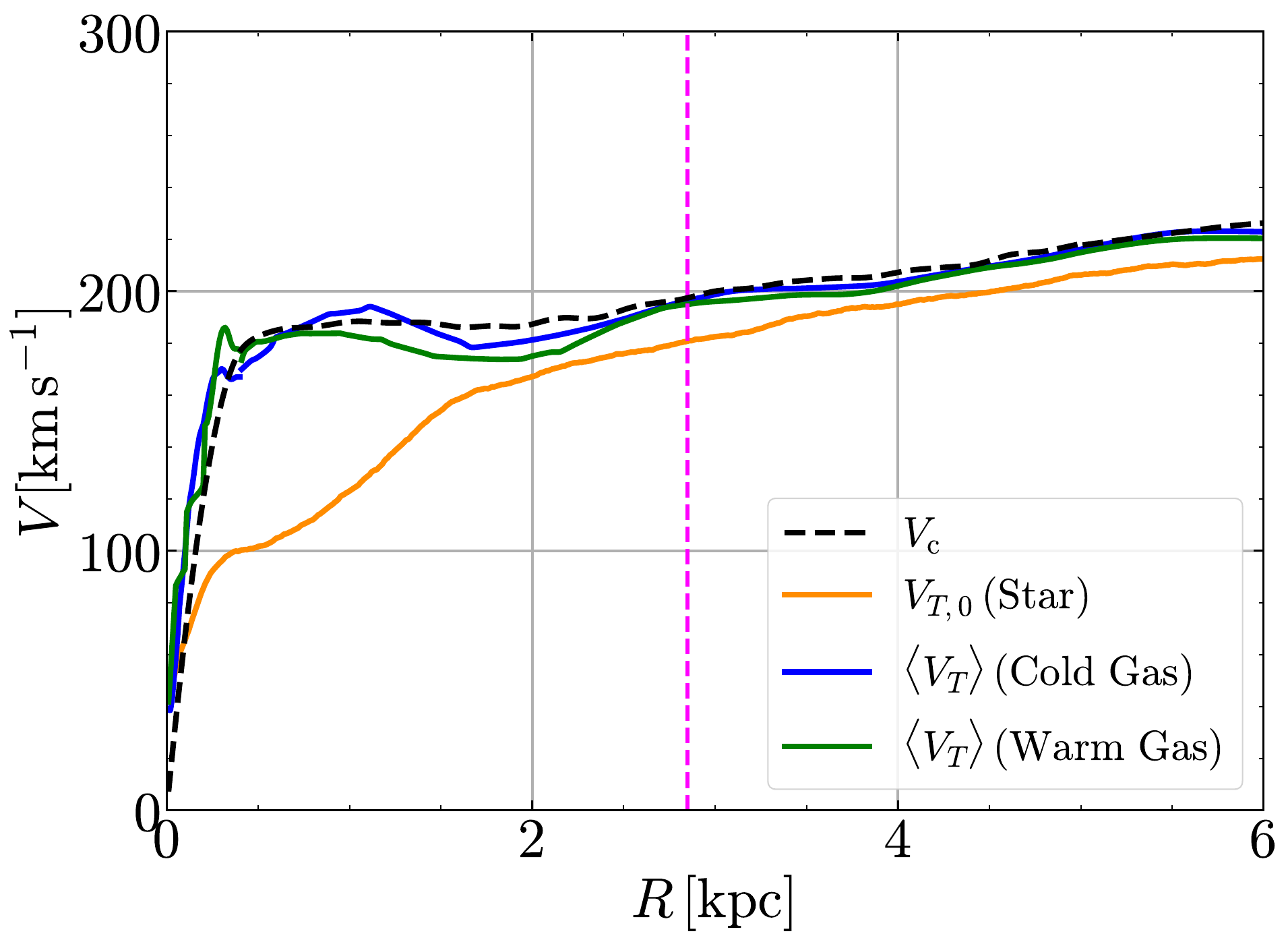}
\caption{
Radial profiles of the circular velocity (black) and the mean rotation velocities of stars (orange), cold gas (blue), and warm gas (green) in model \texttt{C10G10} at $t=2\Gyr$. The vertical dashed line marks the bar length, $R_b$.
}
\label{fig:rot_curve}
\end{figure}

\begin{figure*}[t]
\plotone{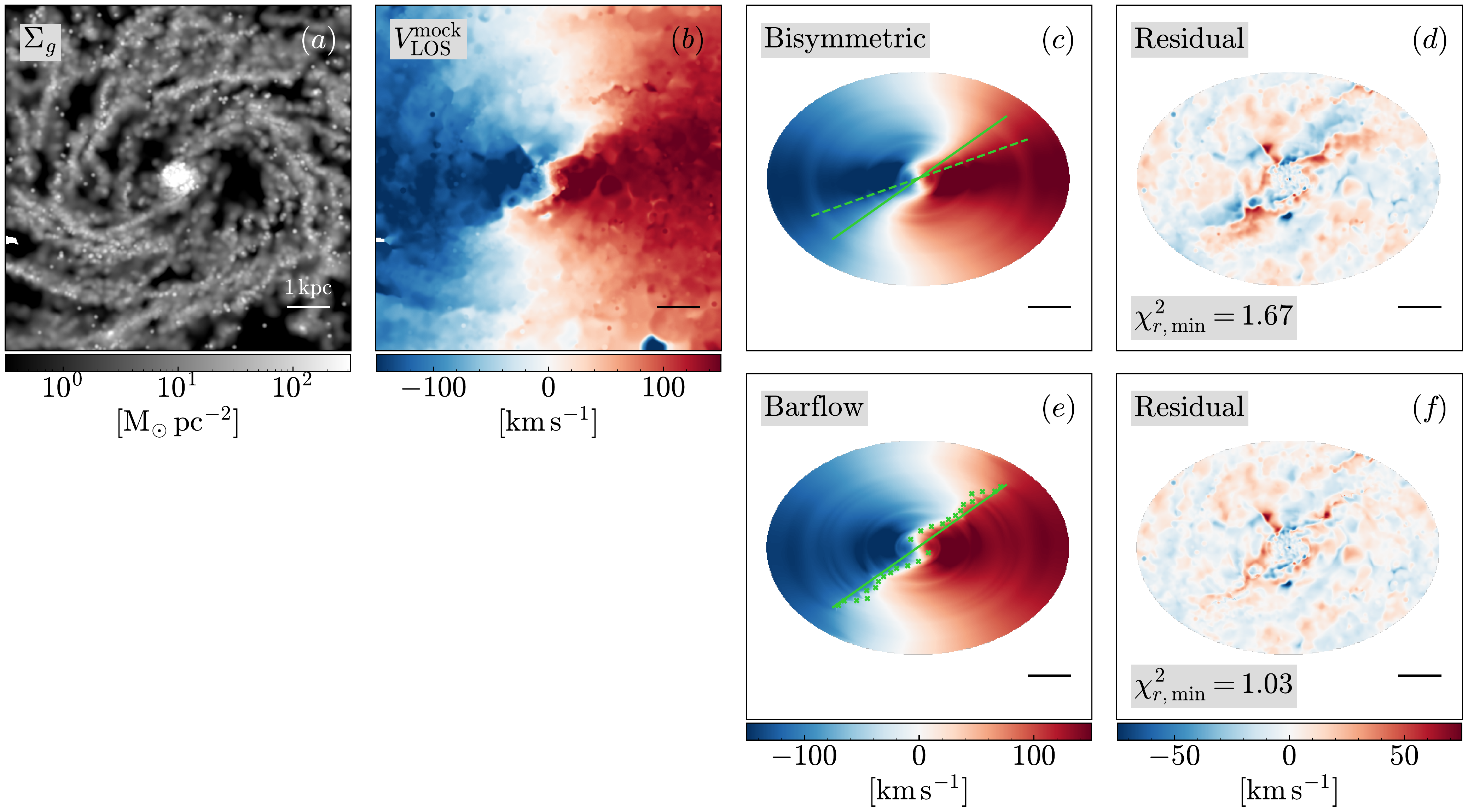}
\caption{Mock \ac{LOS} velocity fitting test for the cold gas component of model \texttt{C10G10} at $t=2 \Gyr$.
$(a)$ Cold-gas surface density map and $(b)$ mock \ac{LOS} velocity field after projection to $i=45^\circ$ and surface density-weighted convolution with a Gaussian \ac{PSF} of \ac{FWHM}$=50\pc$.
$(c)$ Best-fit \ac{LOS} velocity field from the bisymmetric model and $(d)$ its residual map.
$(e)$ Best-fit \ac{LOS} velocity field from the barflow model and $(f)$ its residual map.
In panels $(c)$ and $(e)$, the green solid line shows the true projected stellar bar \ac{PA}.
In panel $(c)$, the green dashed line indicates the projected best-fit global kinematic phase of the bisymmetric model.
In panel $(e)$, the green markers indicate the projected locations of the best-fit shock-front phases $\phi'_s(R)$ at the adopted ring radii.
The horizontal scale bar in each panel spans $1\kpc$.}
\label{fig:mock_LOS_velocity}
\end{figure*}

\Cref{fig:model_comparison_1D} compares the azimuthal profiles of the density-weighted radial and tangential velocities of the cold gas with the best-fit bisymmetric and \textit{barflow} models. The comparison is performed directly in the simulation plane at a fixed radius and is therefore unaffected by \ac{LOS} projection or beam-smearing effects. While the bisymmetric model reproduces the overall sign changes of the velocity field, its sinusoidal form cannot capture the sharp velocity transitions across the shock fronts, particularly near the bar major axis, and consequently underestimates the amplitudes of the bar-driven perturbations. In contrast, the \textit{barflow} model accurately reproduces both the sawtooth-like radial and the archlike tangential velocity profiles. The best-fit shape parameters are $\alpha=0.79$ and $\beta=0.77$, indicating strongly nonsinusoidal gas kinematics. Together with the radially varying shock-front phase, these results demonstrate that the \textit{barflow} model captures the essential features of the simulated gas velocity field.

\subsection{Stellar and Gas Rotation Curves}\label{sec:rotation_curve}

We next examine whether the mean rotation component of the \textit{barflow} model traces the circular velocity derived from the gravitational potential. \Cref{fig:rot_curve} compares the circular velocity, $V_{c}=(R \partial \Phi / \partial R)^{1/2}$ at the disk midplane, with the azimuthally averaged tangential velocities of the stars, cold gas, and warm gas. These mean velocities correspond to $V_{T,0}$ in the bisymmetric model for the stars and $\langle V_T\rangle$ in the \textit{barflow} model for the gas. The stellar rotation curve lies systematically below $V_{c}$, with a larger deficit within the barred region. This offset is consistent with asymmetric drift, in which random stellar motions provide part of the radial support and reduce the mean rotation velocity \citep[e.g.,][]{BT08,Golubov13}. The bar can further enhance random stellar motions and pressure support, increasing the deviation from $V_c$ \citep{Baek25}. Thus, $V_{T,0}$ need not directly trace $V_c$ even when the bisymmetric model accurately describes the mean stellar velocity field.

In contrast, the rotation curves of the cold and warm gases closely follow $V_c$ over most of the radial range. Because gas is collisional and dissipative, random kinetic energy can be converted into heat and radiated away \citep[e.g.,][]{Draine93,Ferriere01,Draine11}. The resulting thermal and turbulent motions \citep[e.g.,][]{Mogotsi16,Kim17} are typically smaller than stellar random motions, therefore provide less radial support. The close agreement between $\langle V_T \rangle$ and $V_c$, however, does not imply nearly circular gas motions. As shown in \autoref{sec:gas_kinematics}, both phases exhibit strong large-scale noncircular motions, and the warm gas also shows feedback-driven velocity fluctuations.

Because the function $g$ has zero azimuthal mean, $\langle V_T\rangle$ separates the mean rotation from the nonsinusoidal velocity perturbations. It provides a good proxy for $V_c$ over most of the disk, including the barred region, whereas the stellar rotation velocity shows a larger systematic deviation from $V_c$.

\section{Discussion}\label{sec:discussion}

In this section, we assess whether the \textit{barflow} model can recover intrinsic gas kinematics from observed \ac{LOS} velocity maps. Using controlled mock observations, we compare its performance with that of the conventional bisymmetric model.

\subsection{Recovery of Intrinsic Gas Kinematics from Line-of-Sight Velocity Maps}
\label{sec:los_vel_fit}

We construct a controlled mock observation from the cold gas component of model \texttt{C10G10} at $t=2\Gyr$, when the galaxy hosts a well-developed stellar bar and prominent bar-driven shocks. The projected \ac{LOS} velocity field is fitted with both the \textit{barflow} model and the conventional bisymmetric model \citep{SS07}. The two models are compared using an identical disk geometry, radial sampling, spatial mask, and fitting procedure, so that any differences in the recovered velocity fields reflect only the adopted velocity parameterization. In the bisymmetric model, the phase of the $m=2$ perturbation is represented by a single global parameter \citep{Sellwood15}, whereas the \textit{barflow} model allows the shock-front phase to vary with radius.

We orient the bar major axis at an angle of $45^\circ$ with respect to the line-of-nodes, project the simulated galaxy to an inclination of $i=45^\circ$, and sample it on a sky-plane grid with $20\pc$ pixels.
The mock \ac{LOS} velocity map is constructed as
\begin{equation}
V_{\rm LOS}^{\rm mock}
=
\frac{
{\rm PSF} * \left(\Sigma_g V_{\rm LOS}\right)
}{
{\rm PSF} * \Sigma_g
},
\end{equation}
where $\Sigma_g$ is the projected gas surface density, $*$ denotes convolution, and the point-spread function (PSF) is a Gaussian with a full width at half maximum (FWHM) of $50\pc$. No random measurement noise is added. The adopted spatial sampling is representative of nearby-galaxy integral field spectroscopy surveys such as PHANGS--MUSE \citep{Bacon10,Emsellem22}.

Assuming that the gas is confined to the disk plane, the projected \ac{LOS} velocity is 
\begin{equation} \label{eq:vlos_projection} 
V_{\rm LOS} = V_{\rm sys} + \sin i \left( V_R\sin\phi + V_T\cos\phi \right), 
\end{equation} 
where $V_{\rm sys}$ is the systemic velocity.
The velocity field is modeled using a tilted-ring approach \citep{Begeman89} with a total of $35$ rings spaced by $100\pc$. The fit includes pixels within the deprojected aperture $R<1.25R_b$. The \textit{barflow} model has six free parameters per ring ($\Delta V_R$, $\langle V_T\rangle$, $\Delta V_T$, $\phi_s$, $\alpha$, and $\beta$), while the bisymmetric model has three ($V_{T,0}$, $V_{R,2}$, and $V_{T,2}$).
Both models also fit the sky-plane coordinates of the kinematic center ($X_c, Y_c$) and the systemic velocity $V_{\rm sys}$, while the bisymmetric model additionally fits a single global kinematic phase $\phi_b$.
The disk \ac{PA} and inclination are fixed to their true values in the mock observation for both models to prevent gas motions from being absorbed into changes in the disk geometry and thereby biasing the recovered velocity components \citep{Schoenmakers97,LopezCoba24a}.
The total numbers of free parameters are therefore $213$ and $109$ for the barflow and bisymmetric models, respectively, as presented in \cref{tab:mock_fit_quality}.

\begin{deluxetable*}{lcccccc}
\tablecaption{
Mock-observation fitting quality and intrinsic velocity recovery
\label{tab:mock_fit_quality}
}
\tablehead{
\colhead{Model} &
\colhead{\(N_{\rm par}\)} &
\colhead{\(\chi_{r,\mathrm{min}}^2\)} &
\colhead{\({\rm BIC}\)} &
\colhead{$\bar{\epsilon}_{\rm rot}$} &
\colhead{$\bar{\epsilon}_{R}$} &
\colhead{$\bar{\epsilon}_{T}$} \\
\colhead{} &
\colhead{} &
\colhead{} &
\colhead{} &
\colhead{$[\kms]$} &
\colhead{$[\kms]$} &
\colhead{$[\kms]$}
}
\startdata
Bisymmetric &
109 &
1.67 &
35974.7 &
61.2 &
57.4 &
31.9 \\
Barflow &
213 &
1.03 &
4356.4 &
10.1 &
23.7 &
19.8
\enddata
\tablecomments{
The reduced chi-squared statistic is defined as $\chi_r^2=\chi^2/(N-N_{\rm par})$, where $N$ and $N_{\rm par}$ are the numbers of fitted pixels and free parameters, respectively. The Bayesian information criterion is computed as ${\rm BIC}=N\ln(\chi^2/N)+N_{\rm par}\ln N$. The quantities $\bar{\epsilon}_{\rm rot}$, $\bar{\epsilon}_R$, and $\bar{\epsilon}_T$ denote the root-mean-square errors of the recovered mean rotational velocity, radial velocity, and tangential velocity perturbation, respectively, averaged over the barred region ($R<R_b$).
}
\end{deluxetable*}

For each trial parameter set, the model velocity field is projected using \cref{eq:vlos_projection}, and the best-fit parameters are obtained by minimizing
\begin{equation}
\chi^2
=
\sum_{n=1}^{N}
\left(
\frac{
V_{{\rm LOS},n}^{\rm mock}
-
V_{{\rm LOS},n}^{\rm model}}
{\sigma_n}
\right)^2,
\label{eq:chi_square}
\end{equation}
where $N$ is the number of fitted pixels. We adopt a uniform velocity scale of $\sigma_n=10\kms$.
We minimize $\chi^2$ in continuous parameter space using \texttt{scipy.optimize.least\_squares} \citep{Virtanen20} which implements the trust-region reflective algorithm \citep{Branch99}.
Because the mock data are noise-free, the resulting $\chi^2$ values are used only for relative model comparison rather than as formal goodness-of-fit statistics.
Parameter uncertainties are estimated using a wild-bootstrap procedure \citep{Efron79,Wu86,Liu88}.

\Cref{fig:mock_LOS_velocity} compares the mock \ac{LOS} velocity field with the best-fit models. Both models reproduce the large-scale projected velocity gradient, but the bisymmetric model leaves coherent residuals around the bar and shock regions, whereas the \textit{barflow} model substantially reduces these residuals. The corresponding fit statistics are summarized in \Cref{tab:mock_fit_quality}. The reduced chi-squared statistic decreases from $\chi^2_{r,\min}=1.67$ for the bisymmetric model to $1.03$ for the \textit{barflow} model. Although the \textit{barflow} model has more free parameters, it also yields a substantially lower \ac{BIC}, indicating that the improved fit is not simply a consequence of increased model flexibility.

The best-fit bisymmetric phase is offset from the true stellar bar position angle by $\Delta\phi_b=18.33^\circ$. Thus, the best-fit phase should be interpreted not as the physical bar orientation but as an effective global $m=2$ phase that minimizes the projected residuals. This illustrates the limitation of describing shock-dominated gas flows with a single fixed phase.

\begin{figure}[t]
\plotone{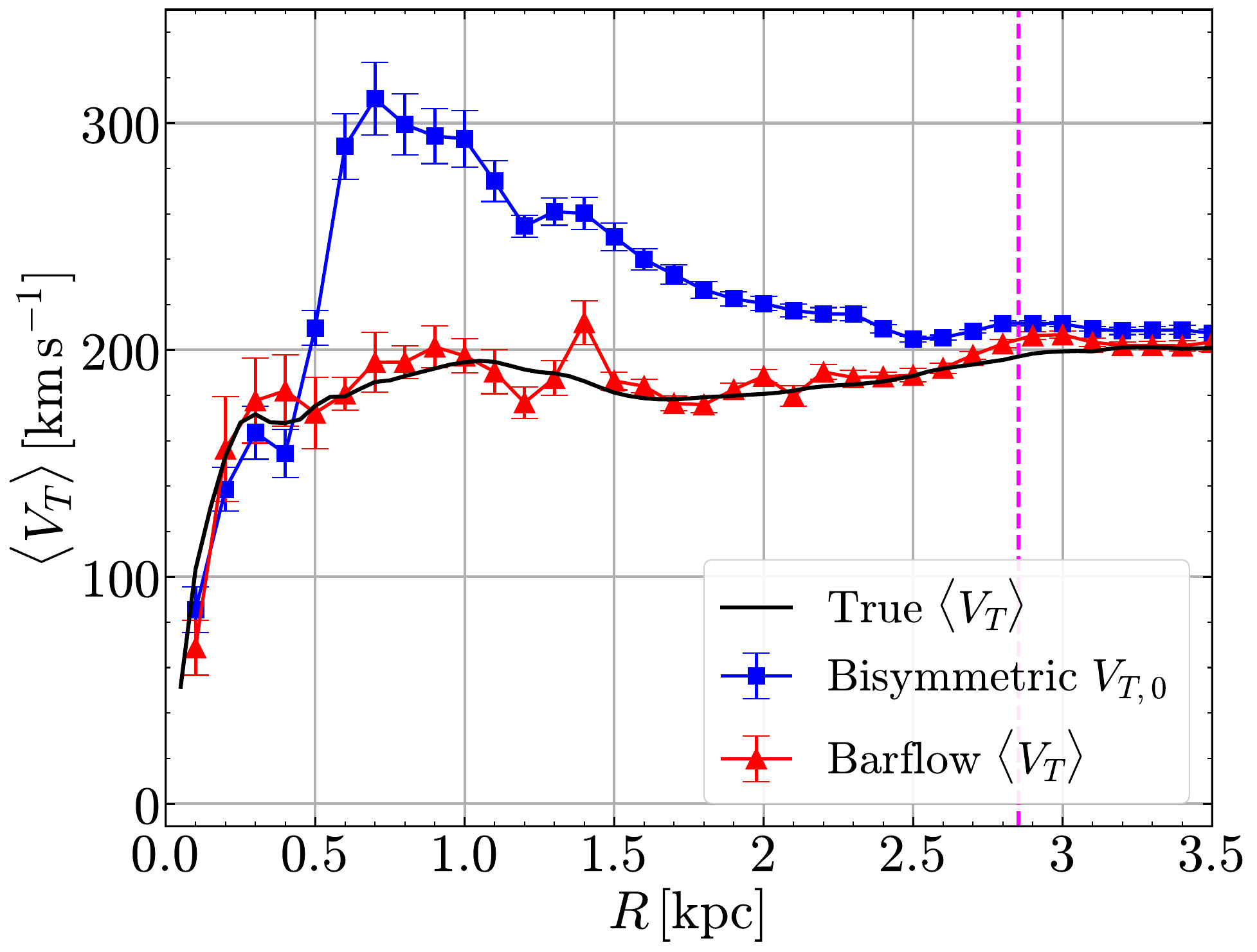}
\caption{Recovery of the rotation curve from the mock \ac{LOS} velocity fitting. The black curve shows the true azimuthally averaged tangential velocity, $\langle V_T\rangle(R)$, measured from the simulation of model \texttt{C10G10} at $t=2\Gyr$. The blue and red curves show the rotation curves recovered by the bisymmetric and \textit{barflow} models, respectively. Error bars indicate bootstrap uncertainties. The vertical dashed line marks the bar length, $R_b$.
}
\label{fig:mock_vrot}
\end{figure}

\begin{figure*}[t]
\centering
\includegraphics[width=0.75\textwidth]{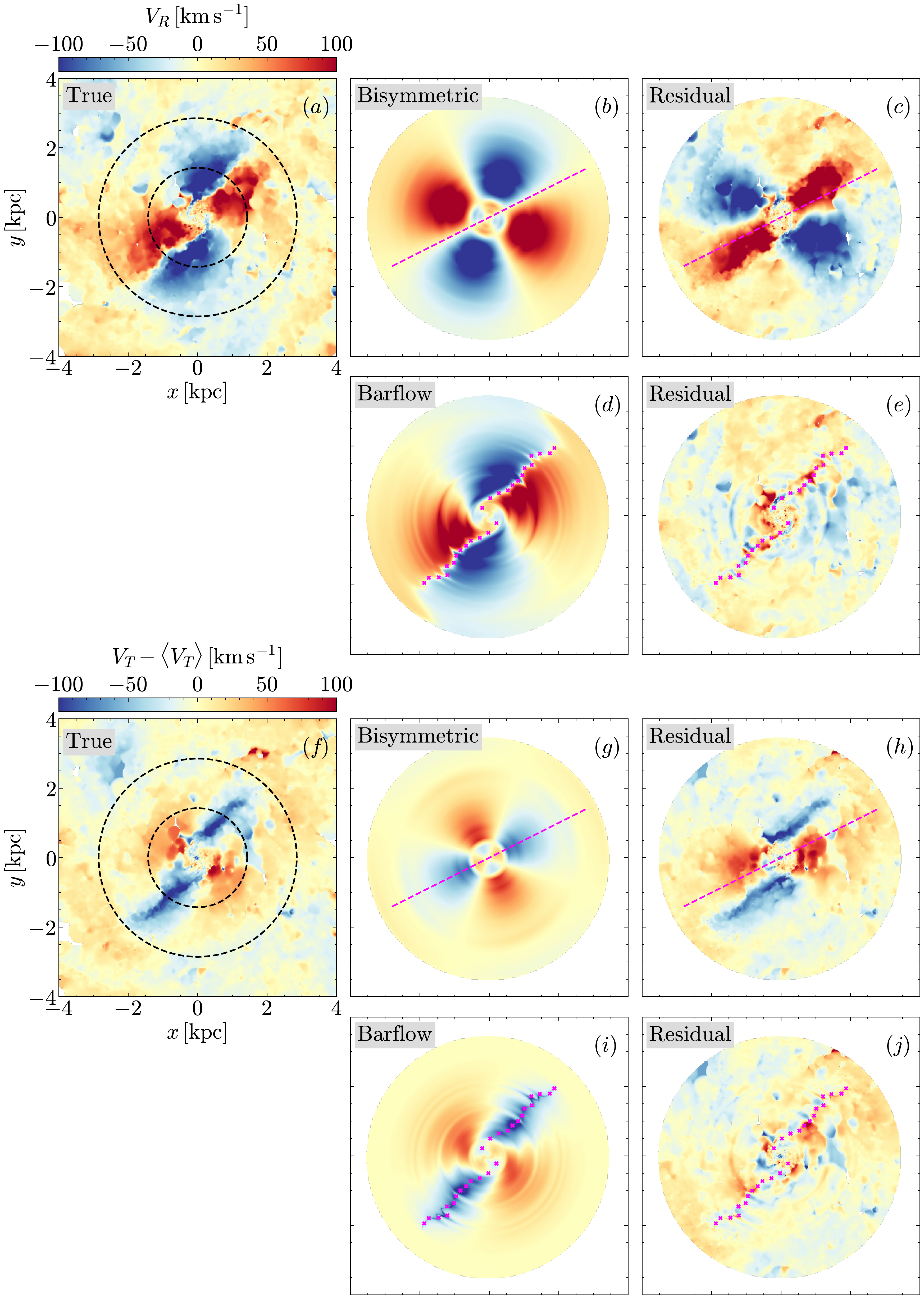}
\caption{Recovery of the intrinsic velocity perturbations from the mock \ac{LOS} velocity field for model \texttt{C10G10} at $t=2\Gyr$.
The upper and lower rows show the radial velocity, $V_R$, and the tangential velocity perturbation, $V_T-\langle V_T\rangle$, respectively. Panels $(a)$ and $(f)$ show the true velocity fields from the simulation. Panels $(b)$ and $(g)$ show the fields recovered by the bisymmetric model, while panels $(d)$ and $(i)$ show those recovered by the \textit{barflow} model. Panels $(c)$, $(e)$, $(h)$, and $(j)$ show the corresponding residuals. The black dashed circles mark $R=0.5R_b$ and $R=R_b$. The magenta dashed lines in the bisymmetric panels indicate the best-fit global kinematic phase, while the magenta markers in the \textit{barflow} panels mark the recovered shock-front phase at the fitted ring radii.
}
\label{fig:mock_2D_velocity}
\end{figure*}

We next examine how well the projected \ac{LOS} fits recover the intrinsic in-plane gas kinematics. \Cref{fig:mock_vrot} compares the recovered rotation curves with the true azimuthally averaged tangential velocity measured directly from the simulation. The \textit{barflow} model closely reproduces the true rotation curve throughout the barred region, whereas the bisymmetric model systematically overestimates it. We quantify the recovery accuracy using the root-mean-square error,
\begin{equation}
\bar{\epsilon}_{\rm rot}
=
\left[
\frac{
\int_{R<R_b}
\left(
\langle V_T\rangle_{\rm recovered}
-
\langle V_T\rangle_{\rm true}
\right)^2
\,dR}
{\int_{R<R_b}dR}
\right]^{1/2},
\label{eq:epsilon_rot}
\end{equation}
yielding $\bar{\epsilon}_{\rm rot}=61.2\kms$ and $10.1\kms$ for the bisymmetric and \textit{barflow} models, respectively.

\begin{figure}[t]
\plotone{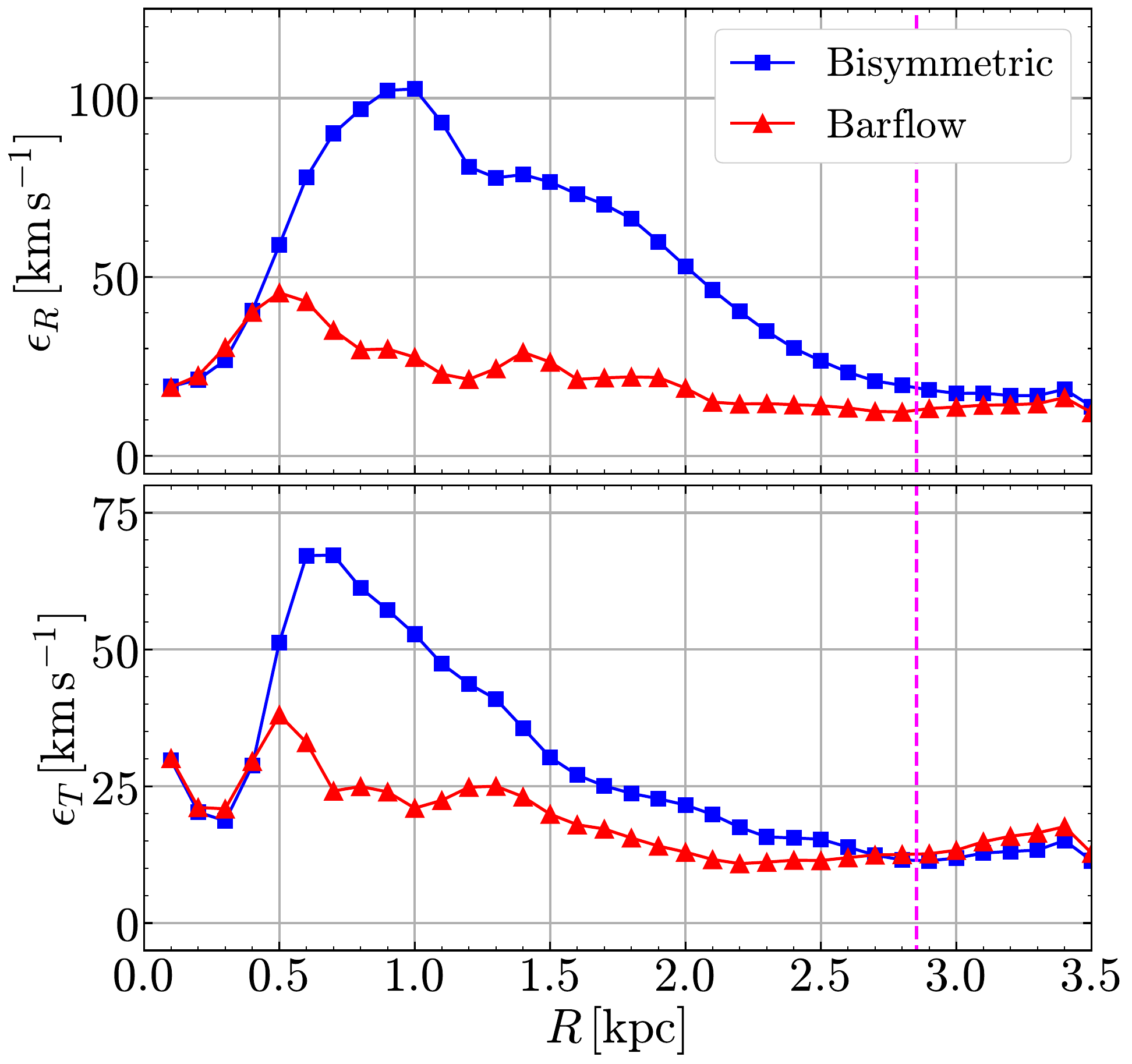}
\caption{Radial distributions of the root-mean-square errors in recovering the intrinsic velocity perturbations for model \texttt{C10G10} at $t=2\Gyr$. The upper and lower panels show the errors in the radial velocity, $V_R$, and the tangential velocity perturbation, $V_T-\langle V_T\rangle$, respectively. The blue and red curves correspond to the bisymmetric and \textit{barflow} models, respectively. The vertical dashed line marks the bar length, $R_b$.}
\label{fig:mock_1D_RMSE}
\end{figure}

\Cref{fig:mock_2D_velocity} compares the recovered intrinsic velocity perturbations. The bisymmetric model imposes a smooth fixed-phase quadrupolar pattern on both $V_R$ and $V_T-\langle V_T\rangle$, reproducing the large-scale velocity structure but failing to capture the sharp velocity transitions associated with the bar-driven shocks. In contrast, the \textit{barflow} model successfully recovers both the radial velocity reversal and the accompanying tangential velocity perturbation along the shock fronts.

The recovery errors of the velocity perturbations are quantified as
\begin{equation}
\epsilon_R(R)
=
\left[
\frac{
\int
\left(
V_{R,\rm recovered}
-
V_{R,\rm true}
\right)^2
\,d\phi}
{\int d\phi}
\right]^{1/2},
\label{eq:RMSE_radial}
\end{equation}
and
\begin{equation}
\epsilon_T(R)
=
\left[
\frac{
\int
\left(
\delta V_{T,\rm recovered}
-
\delta V_{T,\rm true}
\right)^2
\,d\phi}
{\int d\phi}
\right]^{1/2},
\label{eq:RMSE_tangential}
\end{equation}
where $\delta V_T\equiv V_T-\langle V_T\rangle$. The radial distributions of $\epsilon_R$ and $\epsilon_T$ are plotted in \Cref{fig:mock_1D_RMSE}. The \textit{barflow} model yields smaller errors throughout most of the barred region. Averaged over $R<R_b$, the errors are $\bar{\epsilon}_R=23.7\kms$ and $\bar{\epsilon}_T=19.8\kms$, compared to $57.4\kms$ and $31.9\kms$ for the bisymmetric model. These errors are listed in \cref{tab:mock_fit_quality}.

Our results demonstrate that the advantages of the \textit{barflow} model extend beyond providing an improved fit to the projected \ac{LOS} velocity field: it also enables a substantially more accurate recovery of the intrinsic gas kinematics.
These findings motivate revisiting previous applications of the bisymmetric model to gas velocity fields of barred galaxies \citep[e.g.,][]{Sellwood10,Holmes15,LopezCoba22,Zanger24}, as its assumptions may influence the inferred rotation curves and velocity perturbations, with potential implications for the interpretation of bar-driven flows. In a forthcoming study, we will apply the \textit{barflow} model to spatially resolved gas observations of nearby barred galaxies to improve the reconstruction and physical characterization of their intrinsic bar-driven motions.

\subsection{Recovery of Bar-driven Radial Gas Transport}\label{sec:mdot}

The systematic differences in the recovered $V_R$ field directly affect derived quantities such as the bar-driven gas mass transport rate. To assess this effect, we combine the recovered $V_R$ field from each kinematic model with the simulated gas surface density and compare the inferred radial gas transport. This procedure isolates the impact of the adopted kinematic model without introducing additional uncertainties from gas-mass calibration or tracer-to-mass conversion.

\begin{figure}[t]
\plotone{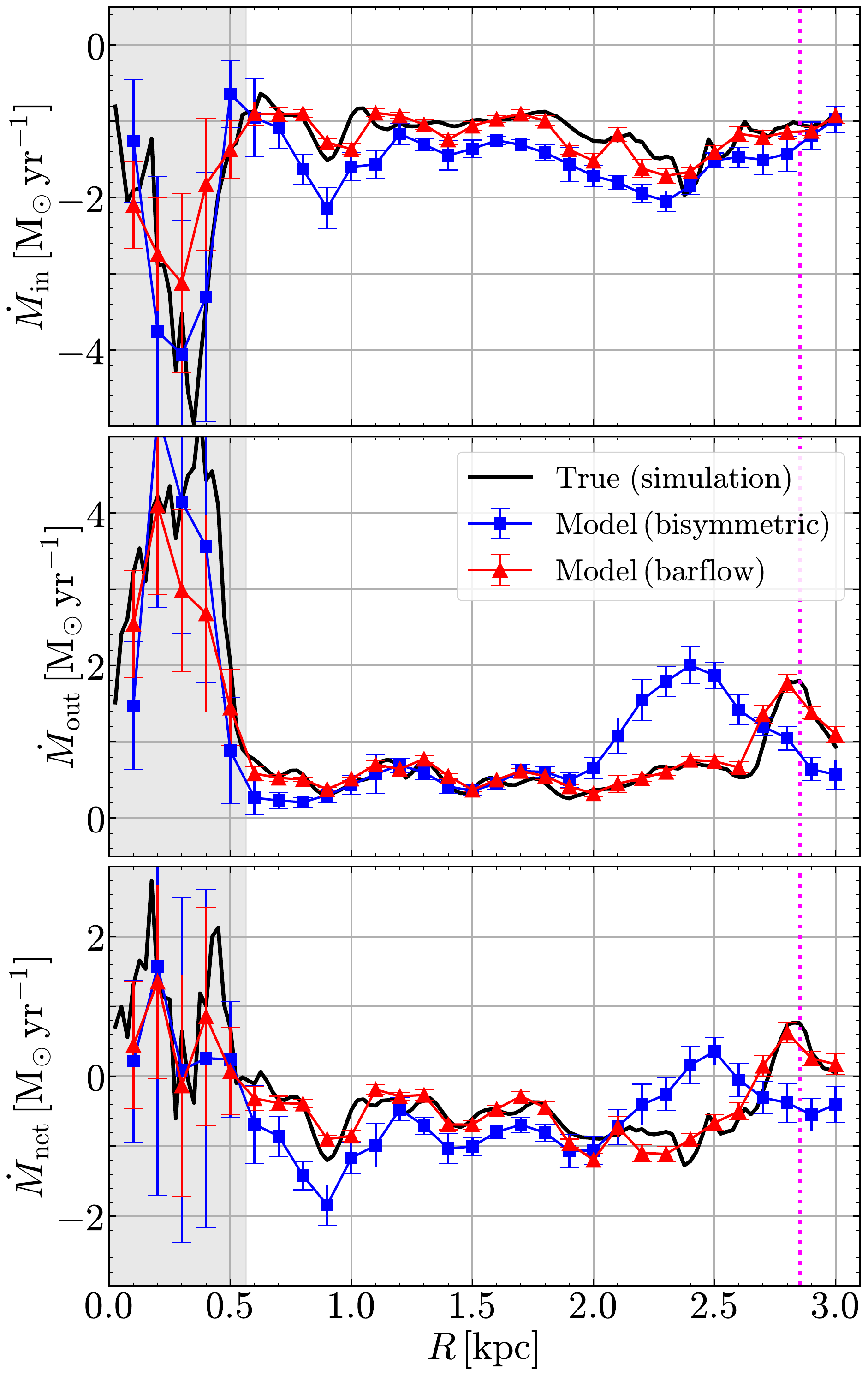}
\caption{Recovered radial mass flux of cold gas from the mock observation of model \texttt{C10G10} at $t=2\Gyr$. From top to bottom, the panels show the inward, outward, and net radial mass fluxes, $\dot{M}_{\rm in}$, $\dot{M}_{\rm out}$, and $\dot{M}_{\rm net}=\dot{M}_{\rm in}+\dot{M}_{\rm out}$, respectively. The black curves show the true fluxes measured directly from the simulation, while the blue and red curves show the fluxes recovered using the bisymmetric and \textit{barflow} models, respectively. Negative values of $\dot{M}_{\rm net}$ correspond to net inflow. The gray-shaded regions on the left indicate the extent of the nuclear gas disk. The vertical dotted line marks the bar length, $R_b$.
}
\label{fig:mock_mdot_cold}
\end{figure}

\begin{figure}[t]
\plotone{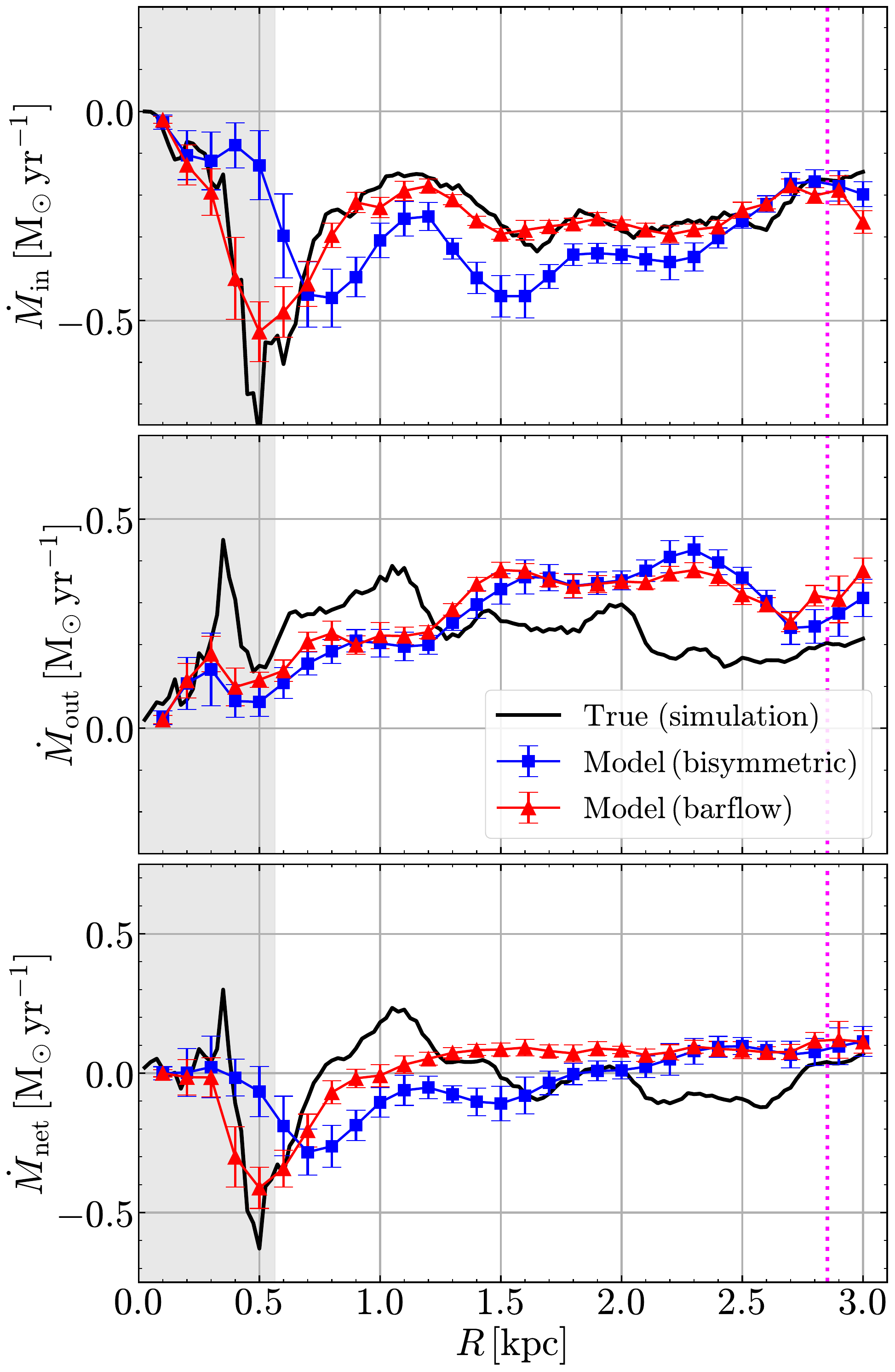}
\caption{
The same as \cref{fig:mock_mdot_cold}, but for the warm gas.
}
\label{fig:mock_mdot_warm}
\end{figure}

Because inward and outward radial motions can coexist at the same radius, we decompose the azimuthally integrated radial gas mass flux into inward, outward, and net components. Adopting the convention that positive $V_R$ denotes outward radial motion, we define
\begin{subequations}\label{eq:mdot}
\begin{align}
\dot{M}_{\rm in}(R)
&=
\int_{V_R<0}
\Sigma_g(R,\phi)V_R(R,\phi)R\,d\phi ,
\label{eq:mdot_in} \\
\dot{M}_{\rm out}(R)
&=
\int_{V_R>0}
\Sigma_g(R,\phi)V_R(R,\phi)R\,d\phi ,
\label{eq:mdot_out}
\end{align}
\end{subequations}
and
\begin{equation}
\dot{M}_{\rm net}(R)
=
\dot{M}_{\rm in}(R)
+
\dot{M}_{\rm out}(R).
\label{eq:mdot_net}
\end{equation}
By definition, $\dot{M}_{\rm in}<0$ and $\dot{M}_{\rm out}>0$, so that negative and positive values of $\dot{M}_{\rm net}$ correspond to net radial inflow and outflow, respectively. 

\Cref{fig:mock_mdot_cold} compares the cold-gas mass fluxes recovered from the two kinematic models with the true fluxes measured directly from model \texttt{C10G10} at $t=2\Gyr$. The \textit{barflow} model reproduces the inward cold-gas flux much more accurately than the bisymmetric model over most of the barred region, and also provides better estimates of the outward and net fluxes. By contrast, the bisymmetric model systematically overestimates the magnitude of the radial mass transport because it fails to recover the shock-associated structure of the intrinsic $V_R$ field. The same analysis for the warm gas is shown in \Cref{fig:mock_mdot_warm}. The \textit{barflow} model again recovers the inward mass flux substantially better than the bisymmetric model, indicating that the large-scale inflow of the warm phase remains governed by the bar-driven flow. However, the outward and net fluxes are reproduced less accurately. This is likely because a significant fraction of the outward-moving warm gas is driven by stellar feedback and turbulent expansion, rather than by the coherent bar-driven streaming described by the \textit{barflow} model.

The comparison between the cold and warm phases illustrates both the strength and the limitation of the \textit{barflow} model. For gas whose radial transport is dominated by bar-driven shocks, the \textit{barflow} model provides a substantially more accurate recovery of intrinsic radial mass transport than the bisymmetric model. However, when outward motions are dominated by stellar feedback and turbulence, $\dot{M}_{\rm out}$ and, consequently, $\dot{M}_{\rm net}$ become less reliable diagnostics of bar-driven gas transport than $\dot{M}_{\rm in}$.

\begin{figure}[t]
\plotone{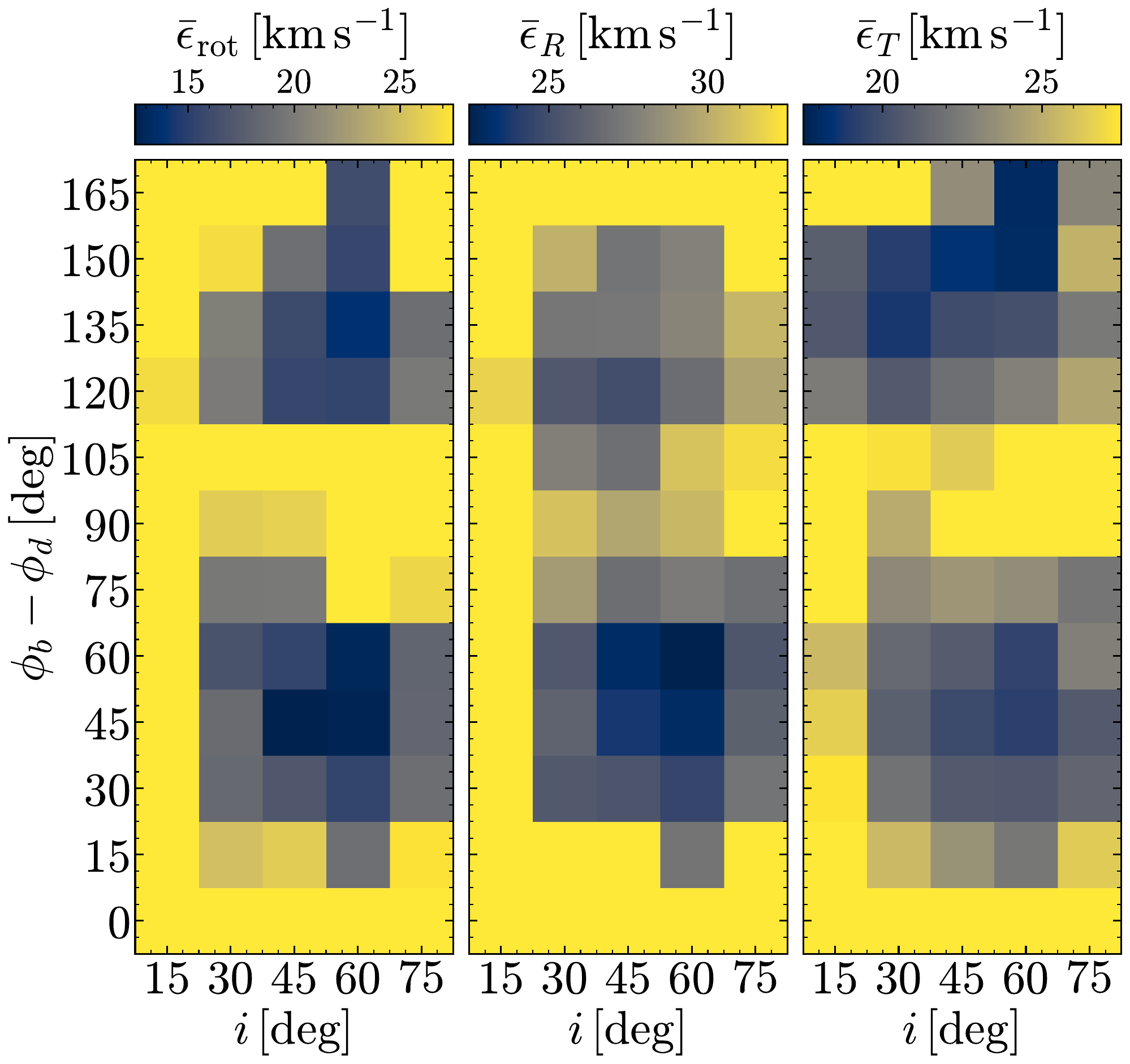}
\caption{Projection-dependent recovery errors of the \textit{barflow} model in the mock \ac{LOS} velocity fitting experiment. The panels show $\bar{\epsilon}_{\rm rot}$, $\bar{\epsilon}_{R}$, and $\bar{\epsilon}_{T}$ as functions of the disk inclination, $i$, and the angle $\phi_b-\phi_d$ between the bar major axis and the line of nodes, measured in the disk plane.}
\label{fig:geometry_test}
\end{figure}

\subsection{Applicability and Limitations of the \textit{Barflow} Model}
\label{subsec:limitations}

The \textit{barflow} model is designed to describe the large-scale response of cold and warm gases to a stellar bar, namely planar, approximately bisymmetric streaming motions with shock-associated velocity transitions. It is therefore not intended to represent all forms of gas motions in barred galaxies. In systems strongly affected by tidal interactions, lopsided gas distributions, extraplanar gas, or feedback-driven outflows, a significant fraction of the observed velocity field may be unrelated to the bar-driven flow \citep[e.g.,][]{Veilleux05,Jog09,Zaritsky13,Kruk17,Leaman19,Watkins23}. In such cases, the fitted shock phase and velocity amplitudes should not be interpreted solely in terms of bar-driven gas dynamics.

The recovery of intrinsic gas motions is also limited by the viewing geometry. To quantify this effect, we repeat the mock \ac{LOS} fitting experiment over a grid of disk inclinations and bar orientations while keeping the intrinsic velocity field and all fitting procedures identical to those adopted in \autoref{sec:los_vel_fit}. For each projection, we measure the recovery errors of the mean rotation, $\bar{\epsilon}_{\rm rot}$, and the radial and tangential velocity perturbations, $\bar{\epsilon}_R$ and $\bar{\epsilon}_T$. The results are displayed in \Cref{fig:geometry_test}. The recovery is most reliable for moderately inclined disks with bars at intermediate in-plane angles to the line of nodes. The errors increase toward face-on and highly inclined systems because of reduced \ac{LOS} sensitivity and stronger projection effects, respectively. They also increase when the bar is nearly parallel or perpendicular to the line of nodes, where the signatures of noncircular motions become increasingly degenerate. These trends follow from the geometry of the \ac{LOS} projection (\cref{eq:vlos_projection}) and are analogous to those encountered in conventional bisymmetric modeling \citep[e.g.,][]{SS07,LopezCoba24a}. 

The \textit{barflow} model is therefore best suited to moderately inclined $(30^{\circ} \lesssim i \lesssim 60^{\circ})$ barred galaxies with sufficient spatial resolution to resolve the shock-induced velocity transition and with the bar neither parallel nor perpendicular to the line of nodes $(30^{\circ} \lesssim \phi_b - \phi_d \lesssim 60^{\circ}; \, 120^{\circ} \lesssim \phi_b - \phi_d \lesssim 150^{\circ})$. For less favorable viewing geometries, the recovered kinematics should be interpreted together with the residual velocity field and independent constraints on the disk geometry, bar properties, and gas or dust-lane morphology.

\section{Summary}\label{sec:summary}

We have presented self-consistent $N$-body/hydrodynamical simulations of isolated, Milky Way-like disk galaxies to investigate how gas regulates stellar bar formation and secular evolution. We also examined the distinct bar-driven kinematic responses of stars and gas and found that the conventional bisymmetric model poorly reproduces shock-associated gas motions. To address this limitation, we introduced a new analytic framework, the \textit{barflow} model, which recovers the intrinsic gas flow substantially more accurately in mock observations (see \cref{fig:model_comparison_1D,fig:mock_2D_velocity}). The \textit{barflow} model thus provides a practical tool for inferring bar-driven gas kinematics from observed velocity fields. Our main results are summarized as follows:

\begin{enumerate}

\item The presence of a dynamically cold gaseous disk delays, and in the most gas-rich model prevents, stellar bar formation over the duration of the simulation. In models with moderate gas fractions ($f_g\leq0.1$), gas promotes higher-order ($m>2$) spiral modes that interfere with the coherent growth of the global bar instability. In the most gas-rich model ($f_g=0.2$), these spirals drive rapid central gas inflow early in the evolution, producing a massive \ac{CMC} that establishes or strengthens an \ac{ILR} and suppresses the wave feedback required for swing amplification, and hence bar formation.

\item Once a bar forms, gas weakens the bar and suppresses its subsequent secular evolution. Bars in disks with higher $f_g$ remain systematically weaker than in the gas-free model, and the bar is eventually destroyed in the $f_g=0.1$ model. This weakening results from continued \ac{CMC} growth driven by bar-induced gas inflow together with positive gravitational torques from the gas, which compensate for the amount of angular momentum transferred to the dark matter halo.
Consequently, the secular slowdown and growth of the bar are both suppressed.

\item Gas also changes the coupled evolution of the bar pattern speed and bar length. Bars in gas-bearing models form with higher initial pattern speeds because of the early spiral-driven growth of the \ac{CMC}. Subsequent gas torques suppress the secular decline of the pattern speed, thereby limiting the outward migration of the corotation radius and the growth of the bar. As a result, bars in the gas-bearing models remain mostly in the fast-bar regime (${\mathcal R}\leq1.4$), whereas the gas-free bar evolves into the slow-bar regime (${\mathcal R}>1.4$).

\item Stars and gas respond differently to the same barred potential. The mean stellar velocity field is well described by a smooth bisymmetric pattern because collisionless orbital mixing smooths the motions of individual stars \citep{Baek25}. In contrast, the cold and warm gases largely follow elongated, $x_1$-like streamlines and retain sharp shock-driven velocity transitions; their kinematic phases vary systematically with radius and follow the curved leading dust lanes. The hot gas, however, is dominated by stellar feedback and thermal pressure, and is largely decoupled from the ordered bar-driven flow. Despite these strong noncircular motions, the mean rotation velocities of the cold and warm gases closely follow the circular velocity derived from the gravitational potential over most of the disk, whereas the mean stellar rotation velocity lies systematically below it, particularly within the barred region.

\item The conventional bisymmetric model cannot reproduce the shock-driven velocity structure of the cold and warm gases. We therefore introduce the \textit{barflow} model, which combines sawtooth-like radial and archlike tangential velocity profiles with a radially varying shock-front phase. In mock \ac{LOS} velocity fitting experiments, the \textit{barflow} model recovers the intrinsic rotation curve, bar-driven velocity perturbations, and radial gas transport substantially more accurately than the conventional bisymmetric model.

\end{enumerate}

\section*{Acknowledgments}
We are grateful to the referee for the constructive and helpful
report. The work of W-T.K. was supported by the grant of the National Research Foundation of Korea (RS-2025-00517264).
T.K. acknowledges support from the Basic Science Research Program through the National Research Foundation of Korea (NRF) funded by the Ministry of Education (RS-2025-25399934).

\bibliography{biblist}{}
\bibliographystyle{aasjournal}


\end{document}